# On Probabilistic Alternating Simulations

Chenyi Zhang and Jun Pang

Faculty of Sciences, Technology and Communication
University of Luxembourg
6, rue Richard Coudenhove-Kalergi, L-1359 Luxembourg

**Abstract.** This paper presents simulation-based relations for probabilistic game structures. The first relation is called probabilistic alternating simulation, and the second called probabilistic alternating forward simulation, following the naming convention of Segala and Lynch. We study these relations with respect to the preservation of properties specified in probabilistic alternating-time temporal logic.

## 1 Introduction

Simulation relations [Mil89] have proved to be useful for comparing the behavior of concurrent systems, which can be formally interpreted as labeled transition systems. The study of logic characterization of simulation is to build its connection to a modal or temporal logic which can be used to formulate some interesting properties. Soundness of logic characterization requires simulation preserve the satisfaction of logic formulas, while completeness shows the relation has the same strength as the logic. Intuitively, the fact that one state $s_1$ simulates another state $s_2$ can be used to establish the relation that any possible behavior of $s_1$ is also possible on $s_2$. Thus it can preserve certain desirable properties[1] formulated in temporal logics like CTL [Eme90]. Simulation relations have set up the foundations for constructing correct abstractions.

*Related work.* Segala and Lynch [SL95] extend the classical notions of simulation for probabilistic automata [Seg95b], a general extension of labeled transition systems which admits both probabilistic and nondeterministic behaviors. Their main idea is to relate probability distributions over states, instead of relating individual states. They show soundness of the logical characterization of probabilistic simulation, which preserves probabilistic CTL formulas [Han94] without negation and existential quantification. Segala introduces the notion of probabilistic forward simulation, which relates states to probability distributions over states and is sound and complete for trace distribution precongruence [Seg95a,LSV07]. Logic characterization of strong and weak probabilistic bisimulation has been studied in [DGJP02,PS07].

Alur, Henzinger and Kupferman [AHK97,AHK02] define ATL (alternating-time temporal logic) to generalize CTL for game structures by requiring each path quantifier to be parametrized with a set of agents. Game structures are more general than LTS, in the sense that they allow both collaborative and adversarial behaviors of individual agents in a system, and ATL can be used to express properties like "a set of agents can enforce a specific outcome of the system". Alternating refinement relations, in particular alternating simulation, are introduced later in [AHKV98]. Alternating simulation is a natural game-theoretic interpretation of the classical simulation in two-player games. Logic characterization of this simulation concentrates on a subset of ATL$^\star$ formulas where negations are only allowed at proposition level and all path quantifiers are parametrized by a prefixed set of agents $A$. This sublogic of ATL$^\star$ contains all formulas expressing the properties the agents in $A$ can enforce no matter what the other agents do. Alur et al. [AHKV98] have proved both soundness and completeness of the logic characterization.

---
[1] For example, safety properties stating "nothing bad can happen".

*Our contribution.* In this apper, we introduce two notions of simulation for probabilistic game structures – probabilistic alternating simulation and forward simulation, following the aforementioned results [Seg95a,SL95,AHKV98]. We prove the soundness of logical characterization of probabilistic alternating simulation relations, by showing that they preserve a fragment of a probabilistic extension of ATL.

*Structure of the paper.* The rest of the paper is organized as follows. We briefly explain some basic notations that are used throughout the paper in Sect. 2. Sect. 3 introduces the notion of probabilistic game structures and the definition of probabilistic executions. In Sect. 4 we present PATL an extension of the alternating-time temporal logic [AHK02] for probabilistic systems, and roughly discuss its model checking problem. We define probabilistic alternating simulation and forward simulation in Sect. 5, and show their soundness for preserving properties specified in PATL in Sect. 6. Probabilistic alternating bisimulation is shortly discussed in Sect. 7. We conclude the paper with some future research topics in Sect. 8.

## 2 Preliminaries

This section contains basic notions that are used in the technical part. Let $S$ be a set. A discrete probabilistic distribution $\Delta$ over $S$ is a function of type $S \to [0,1]$, satisfying $\sum_{s \in S} \Delta(s) = 1$. We write $\mathcal{D}(S)$ for the set of all such distributions. For a set $S' \subseteq S$, define $\Delta(S') = \sum_{s \in S'} \Delta(s)$. Given two distributions $\Delta_1, \Delta_2$ and $p \in [0,1]$, $\Delta_1 \oplus_p \Delta_2$ is a function of type $S \to [0,1]$ defined as $\Delta_1 \oplus_p \Delta_2(s) = p \cdot \Delta_1(s) + (1-p) \cdot \Delta_2(s)$ for all $s \in S$. Obviously, $\Delta_1 \oplus_p \Delta_2$ is also a distribution. We further extend this notion by combining a set of distributions $\{\Delta_i\}_{i \in I}$ ordered by an indexed set $\{p_i\}_{i \in I}$ into a distribution $\sum_{i \in I} p_i \Delta_i$, where $p_i \in [0,1]$ for all $i \in I$ and $\sum_{i \in I} p_i = 1$. $\overline{s}$ is called a point distribution satisfying $\overline{s}(s) = 1$ and $\overline{s}(t) = 0$ for all $t \neq s$. Let $\Delta \in \mathcal{D}(S)$, write $\lceil \Delta \rceil$ for the *support* of $\Delta$ as the set $\{s \in S \mid \Delta(s) > 0\}$.

Let $S = S_1 \times S_2 \times \cdots \times S_n$, then $\boldsymbol{s} \in S$ is a vector of length $n$. We may also write $\boldsymbol{s} = \langle s_1, s_2, \ldots, s_n \rangle$, with $\boldsymbol{s}(i) = s_i \in S_i$. Given a finite sequence $\alpha = s_1 s_2 \ldots s_n \in S^*$, write $last(\alpha)$ for $s_n$. Let $S' \subseteq S$, then $\alpha \mid S'$ is a subsequence of $\alpha$ with exactly the elements not in $S'$ removed. Given $L \subseteq S^*$, write $L \mid S'$ for the set $\{(\alpha \mid S') \mid \alpha \in L\}$.

## 3 Probabilistic Game Structures

Assume a set of players $\Sigma = \{1, 2, \ldots, \mathtt{k}\}$. A probabilistic game structure (PGS) $\mathcal{G}$ is defined as a tuple $\langle S, s_0, \mathcal{L}, \mathtt{Act}, \delta \rangle$, where

- $S$ is a finite set of states, with $s_0$ the initial state,
- $\mathtt{Act} = \mathtt{Act}_1 \times \mathtt{Act}_2 \times \cdots \times \mathtt{Act}_\mathtt{k}$ is a set of joint actions, where $\mathtt{Act}_i$ is the set of actions for player $i = 1, \ldots, \mathtt{k}$,
- $\mathcal{L} : S \to 2^{\mathtt{Prop}}$ is the labelling function,
- $\delta : S \times \mathtt{Act} \to \mathcal{D}(S)$ is a transition function.

A play $\rho$ is a (finite or infinite) sequence $s_0 \boldsymbol{a}_1 s_1 \boldsymbol{a}_2 s_2 \ldots$, such that $\boldsymbol{a}_i \in \mathtt{Act}$ and $\delta(s_{i-1}, \boldsymbol{a}_i)(s_i) > 0$ for all $i$. Write $|\rho|$ for the length of a run $\rho$, which is the number of transitions in $\rho$, and $|\rho| = \infty$ if $\rho$ is infinite. We write $\rho(i)$ for the $i$-th state in $\rho$ starting from 0, and $\rho[i, j]$ for the subsequence starting from $i$-th state and ending at the $j$-th state, provided $0 \le i \le j \le |\rho|$. Note that the players choose their next moves simultaneously, but their moves may or may not be cooperative. If on state $s$ each player $i$ performs action $a_i$, then $\delta(s, \langle a_1, a_2, \ldots a_k \rangle)$ is the distribution for the next reachable states. In the following discussion, we fix a probabilistic game structure $\mathcal{G}$.

We assume that the transition relation is total on the set $\mathtt{Act}$. Note that this does not pose any limitation on the expressiveness of the model. If an action $c \in \mathtt{Act}_i$ of player $i$ is not supposed to be enabled on state $s$ for player $i$, we may find another action $c' \in \mathtt{Act}_i$ and define $c$ to have the same effect as $c'$ on $s$. Since player $i$ knows the current state, he also knows the set of actions available to him, so that as a rational player he will not choose actions that are not enabled. This allows such models to express systems in which on some states the available (joint) actions are proper subsets of $\mathtt{Act}$.[2] We may even disable a particular player on a state. A player $i$ is disabled on $s$ if $\delta(s, \boldsymbol{a}) = \delta(s, \boldsymbol{a}')$ for all action vectors $\boldsymbol{a}, \boldsymbol{a}' \in \mathtt{Act}$ satisfying $\boldsymbol{a}(j) = \boldsymbol{a}'(j)$ for all $j \ne i$. A PGS is *turn-based* if all but one player is disabled on $s$ for all $s \in S$. A probabilistic game structure can be regarded as a generalization of a concurrent game structure of [AHK02]. From a state $s \in S$, each player $i$ may choose an action from $\mathtt{Act}_i$ and together they resolve the nondeterminism. On the other hand, a PGS is more stratified on external actions than some of the existing models.[3]

---

[2] In the literature some authors encode available actions for player $i$ as a function of type $S \to 2^{\mathtt{Act}_i} \setminus \{\emptyset\}$.

[3] For example, a one-player PGS resembles a reactive system of [vGSS95], and a two-player turn-based PGS (assuming they alternately act) loosely simulates a simple probabilistic automaton [Seg95b], in the way that one player performs external actions and the other resolves nondeterminism after the previous player's move is done.

A strategy of a player $i \in \Sigma$ is a function of type $S^+ \to \mathcal{D}(\texttt{Act}_i)$. We write $\Pi_i^{\mathcal{G}}$ for the set of strategies of player $i$ in $\mathcal{G}$.[4] A play $\rho$ is compatible with an $i$-strategy $\pi_i$, if $\boldsymbol{a}_k(i) \in \lceil \pi_i(\rho[0, k-1]|S)\rceil$ for all $k \leq |\rho|$. Given a vector of strategies $\boldsymbol{\pi} \in \Pi_1^{\mathcal{G}} \times \Pi_2^{\mathcal{G}} \times \cdots \times \Pi_{|\Sigma|}^{\mathcal{G}}$, a run $\rho$ is compatible with $\boldsymbol{\pi}$ if $\boldsymbol{a}_k(i) \in \lceil \boldsymbol{\pi}(i)(\rho[0, k-1]|S)\rceil$ for all $k \leq |\rho|$ and $i = 1, \ldots, \texttt{k}$. Write $\mathcal{G}(\boldsymbol{\pi}, s)$ for the set of infinite plays compatible with every strategy in $\boldsymbol{\pi}$ starting from $s \in S$, and $\mathcal{G}^*(\boldsymbol{\pi}, s)$ the set of finite plays in $\mathcal{G}$ that are compatible with $\boldsymbol{\pi}$ starting from $s$.

The set of finite plays compatible to a strategy vector $\boldsymbol{\pi}$ is also called a set of *cones* [Seg95b], with each finite play $\alpha$ representing the set of infinite plays prefixed by $\alpha$. Given a state $s_0 \in S$, we can derive the probability for every member in $S^+$ compatible with $\boldsymbol{\pi}$, by recursively defining a function $Pr_{\mathcal{G}(\boldsymbol{\pi}, s_0)}$ from $S^+$ to $[0, 1]$ as follows. This function $Pr_{\mathcal{G}(\boldsymbol{\pi}, s_0)}$ can be further generalized as the probability measure to the $\sigma$-field $\mathcal{F}_{\mathcal{G}, \boldsymbol{\pi}, s_0} \subseteq \mathcal{G}(\boldsymbol{\pi}, s_0)$ which is a unique extension from the set of cones $\mathcal{G}^*(\boldsymbol{\pi}, s)$ closed by countable union and complementation, in a way similar to [Seg95b]:

- $Pr_{\mathcal{G}(\boldsymbol{\pi}, s_0)}(s_0) = 1$,
- $Pr_{\mathcal{G}(\boldsymbol{\pi}, s_0)}(\alpha \cdot s) = Pr_{\mathcal{G}(\boldsymbol{\pi}, s_0)}(\alpha) \cdot \overline{\delta}(last(\alpha), \langle \boldsymbol{\pi}(1)(\alpha), \boldsymbol{\pi}(2)(\alpha), \ldots, \boldsymbol{\pi}(\texttt{k})(\alpha)\rangle)(s)$,

where $\overline{\delta}(s, \langle \Delta_1, \Delta_2, \ldots, \Delta_\texttt{k}\rangle)$ is a distribution over states derived from $\delta$ and the vector of action distributions defined by

$$\overline{\delta}(s, \langle \Delta_1, \ldots, \Delta_\texttt{k}\rangle) = \sum_{i \in \{1, \ldots, \texttt{k}\}, a_i \in \lceil \Delta_i \rceil} \Delta_1(a_1) \cdot \ldots \cdot \Delta_\texttt{k}(a_\texttt{k}) \cdot \delta(s, \langle a_1, \ldots, a_\texttt{k}\rangle).$$

Given $A \subseteq \Sigma$, sometimes we write $\boldsymbol{\pi}(A)$ for a vector of $|A|$ strategies $\{\pi_i\}_{i \in A}$, and $\Pi(A)$ for the set of all such strategy vectors. Write $\overline{A}$ for $\Sigma \setminus A$. Given $A \cap A' = \emptyset$, strategy vectors $\boldsymbol{\pi} \in \Pi(A)$ and $\boldsymbol{\pi}' \in \Pi(A')$, $\boldsymbol{\pi} \cup \boldsymbol{\pi}'$ is the vector of strategies $\{\pi_i\}_{i \in A} \cup \{\pi'_j\}_{j \in A'}$ that combines $\boldsymbol{\pi}$ and $\boldsymbol{\pi}'$.

We also define strategies of *finite depth* by restricting the size of their domains, by writing $\pi \in \Pi_i^{\mathcal{G}, n}$ as a *level-n* strategy, i.e., $\pi$ is a function from traces of states with length up to $n$ (i.e., the set $\bigcup_{m \in \{1, 2, \ldots, n\}} S^m$) to $\mathcal{D}(\texttt{Act}_i)$. Given a set of strategies $\{\pi_i\}_{i \in I}$ of the same domain, and $\{p_i\}_{i \in I}$ with $\sum_{i \in I} p_i = 1$, let $\pi = \sum_{i \in I} p_i \cdot \pi_i$ be a (combined) strategy, by letting $\pi(\gamma) = \sum_{i \in I} p_i \cdot \pi_i(\gamma)$ for all $\gamma$ in the domain.

We overload the function $\overline{\delta}$ as from a state in $S$ and a vector of strategies (of any depth $n$) $\boldsymbol{\pi} \in \Pi_1^{\mathcal{G}, n} \times \Pi_2^{\mathcal{G}, n} \times \cdots \times \Pi_{|\Sigma|}^{\mathcal{G}, n}$ to $\mathcal{D}(S)$, by $\overline{\delta}(s, \boldsymbol{\pi}) = \overline{\delta}(s, \boldsymbol{a})$, where $\boldsymbol{a}(i) = \boldsymbol{\pi}(i)(s)$ for all $i \in \Sigma$. Note each $\boldsymbol{a}(i)$ is a distribution over $\texttt{Act}_i$. We further lift $\overline{\delta}$ to be a transition function from state distributions and strategy vectors to state distributions, by

$$\overline{\delta}(\Delta, \boldsymbol{\pi}) = \sum_{s \in \lceil \Delta \rceil} \Delta(s) \cdot \overline{\delta}(s, \boldsymbol{\pi})$$

---

[4] Sometimes we omit $\mathcal{G}$, if it is clear from the context.

**Probabilistic Executions**

We settle the nondeterminism in a probabilistic game structure by fixing the behaviours of all players represented as strategies. Let $\mathcal{G} = \langle S, s_0, \mathcal{L}, \texttt{Act}, \delta \rangle$ be a PGS, define a *probabilistic execution* $\mathcal{E}$ as in the form of $\langle E, \Delta, \mathcal{L}^{\mathcal{E}}, \delta^{\mathcal{E}} \rangle$, where

- $E \subseteq S^+$ is the set of finite plays starting form a state in the initial distribution and compatible with $\delta^{\mathcal{E}}$, i.e., $s_0 s_1 \ldots s_n \in E$ if $s_0 \in \lceil \Delta \rceil$, and $\delta^{\mathcal{E}}(s_0 \ldots s_i)(s_0 \ldots s_{i+1}) > 0$ for all $0 \leq i < n$,
- $\Delta \in \mathcal{D}(S)$ an (initial) distribution,
- $\mathcal{L}^{\mathcal{E}}$ is the labelling function defined as $\mathcal{L}^{\mathcal{E}}(e) = \mathcal{L}(last(e))$ for all $e \in E$,
- $\delta^{\mathcal{E}} : E \to \mathcal{D}(E)$ is a (deterministic) transition relation, satisfying for all $e \in E$ there exists a (level 1) strategy vector $\boldsymbol{\pi}_e$, such that $\delta^{\mathcal{E}}(e)(e \cdot t) = \overline{\delta}(last(e), \boldsymbol{\pi}_e)(t)$ if $t \in \lceil \overline{\delta}(last(e), \boldsymbol{\pi}_e) \rceil$, and 0 otherwise.

A probabilistic execution of $\mathcal{G}$ can be uniquely determined by a strategy vector $\boldsymbol{\pi}$ starting from a state distribution. Given $\Delta \in \mathcal{D}(S)$, define $\mathcal{E}(\mathcal{G}, \boldsymbol{\pi}, \Delta)$ as the probabilistic execution $\langle E^{\boldsymbol{\pi}}, \Delta, \mathcal{L}^{\boldsymbol{\pi}}, \delta^{\boldsymbol{\pi}} \rangle$, with $E^{\boldsymbol{\pi}} = \bigcup_{s \in \lceil \Delta \rceil} \mathcal{G}^*(\boldsymbol{\pi}, s) \mid S$ for the set of compatible finite plays, $\mathcal{L}^{\boldsymbol{\pi}}$ defined as $\mathcal{L}^{\boldsymbol{\pi}}(e) = \mathcal{L}(last(e))$ for all $e \in E^{\boldsymbol{\pi}}$, and $\delta^{\boldsymbol{\pi}}(e) = \overline{\delta}(last(e), \boldsymbol{\pi}_e)$ for all $e \in E^{\boldsymbol{\pi}}$, where $\boldsymbol{\pi}_e(i) = \boldsymbol{\pi}(i)(e)$ for all $i \in \Sigma$. Intuitively, a probabilistic execution resembles the notion of the same name proposed by Segala and Lynch [Seg95b,SL95], and in this case the strategies of the players altogether represent a single adversary of Segala and Lynch.

## 4 Probabilistic Alternating-Time Temporal Logic

In this section we introduce a probabilistic version of alternating-time temporal logic [AHK02], which focuses on the players ability to enforce a property with an expected probability. Let Prop be a nonempty set of propositions. Probabilistic alternating-time temporal logic (PATL) formulas [CL07] are defined as follows.

$$\phi := p \mid \neg \phi \mid \phi_1 \wedge \phi_2 \mid \langle\!\langle A \rangle\!\rangle^{\bowtie \alpha} \psi$$
$$\psi := \bigcirc \phi \mid \phi_1 \mathtt{U}^{\leq k} \phi_2$$

where $A \subseteq \Sigma$ is a set of players, $\bowtie \in \{<, >, \leq, \geq\}$, $k \in \mathbb{N} \cup \{\infty\}$, $p \in \texttt{Prop}$, and $\alpha \in [0, 1]$. We also write $\psi_1 \mathtt{U} \psi_2$ for $\psi_1 \mathtt{U}^{\leq \infty} \psi_2$ as 'unbounded until'. The symbols $\phi, \phi_1, \phi_2$ are state formulas, and $\psi$ is a path formula. We omit the syntactic sugars in our definition, such as *true* $\equiv p \vee \neg p$ and *false* $\equiv p \wedge \neg p$ for some $p \in \texttt{Prop}$, $\phi_1 \vee \phi_2 \equiv \neg(\neg \phi_1 \wedge \neg \phi_2)$ for state formulas. The path modality R can be expressed by U without introducing negations into path formulas, as we will show later in this section. One may also define $\square^{\leq k} \psi \equiv \textit{false } \mathtt{R}^{\leq k} \psi$, and $\lozenge^{\leq k} \psi \equiv \textit{true } \mathtt{U}^{\leq k} \psi$, where $k \in \mathbb{N} \cup \{\infty\}$. The set of PATL formulas $\mathbb{L}$ are the set of state formulas as defined above. We have the semantics of the path formulas and the state formulas defined as follows.

- $\rho \models \phi$ iff $\mathcal{G}, \rho(0) \models \phi$ where $\phi$ is a state formula,
- $\rho \models \bigcirc \phi$ iff $\rho(1) \models \phi$,

- $\rho \models \phi_1 \mathtt{U}^{\leq k} \phi_2$ iff there exists $i \leq k$ such that $\rho(j) \models \phi_1$ for all $0 \leq j < i$ and $\rho(i) \models \phi_2$,
- $\mathcal{G}, s \models p$ iff $p \in \mathcal{L}(s)$,
- $\mathcal{G}, s \models \neg \phi$ iff $\mathcal{G}, s \not\models \phi$,
- $\mathcal{G}, s \models \phi_1 \wedge \phi_2$ iff $\mathcal{G}, s \models \phi_1$ and $\mathcal{G}, s \models \phi_2$,
- $\mathcal{G}, s \models \langle\!\langle A \rangle\!\rangle^{\bowtie \alpha} \psi$ iff there exists a vector of strategies $\boldsymbol{\pi} \in \Pi(A)$, such that for all vectors of strategies $\boldsymbol{\pi}' \in \Pi(\overline{A})$ for players in $\overline{A}$, we have $Pr_{\mathcal{G}(\boldsymbol{\pi} \cup \boldsymbol{\pi}', s)}(\{\rho \in \mathcal{G}(\boldsymbol{\pi} \cup \boldsymbol{\pi}', s) \mid \rho \models \psi\}) \bowtie \alpha$, where $\rho$ is an infinite play in $\mathcal{G}$, $\alpha \in [0, 1]$, $\phi$, $\phi_1$, $\phi_2$ are state formulas, and $\psi$ is a path formula. Equivalently, given $S$ the state space of a probabilistic game structure $\mathcal{G}$, we write $[\![\phi]\!]$ for $\{s \in S \mid s \models \phi\}$ for all PATL (state) formulas $\phi$. For $\Delta \in \mathcal{D}(S)$, we write $\Delta \models \phi$ iff $\lceil \Delta \rceil \subseteq [\![\phi]\!]$. Intuitively, $\mathcal{G}, s \models \langle\!\langle A \rangle\!\rangle^{\geq \alpha} \psi$ describes the ability of players in $A$ to cooperatively enforce $\psi$ with probability at least $\alpha$ in $s$.

The following lemma is directly from the PATL semantics. If a group of users $A$ can enforce a linear-time temporal logic formula $\psi$ to hold with probability at least $\alpha$ with strategies $\boldsymbol{\pi} \in \Pi(A)$, then at the same time $\boldsymbol{\pi}$ enforces the formula $\neg \psi$ to hold with probability at most $1 - \alpha$. To simplify the notation, we let '$\sim$' denote changes on directions of the symbols in $\{<, >, \leq, \geq\}$, e.g., symbol $\widetilde{\geq}$ for $\leq$, $\widetilde{\leq}$ for $\geq$, $\widetilde{>}$ for $<$, and $\widetilde{<}$ for $>$.

**Lemma 1.** $\mathcal{G}, s \models \langle\!\langle A \rangle\!\rangle^{\bowtie \alpha} \psi$ iff $\mathcal{G}, s \models \langle\!\langle A \rangle\!\rangle^{\widetilde{\bowtie} 1 - \alpha} \neg \psi$

*Proof.* (sketch) For all $\boldsymbol{\pi} \in \Pi(A)$ and $\boldsymbol{\pi}' \in \Pi(\overline{A})$, $s \in S$ and $\psi$ a path formula, we have $Pr_{\mathcal{G}(\boldsymbol{\pi} \cup \boldsymbol{\pi}', s)}(\{\rho \in \mathcal{G}(\boldsymbol{\pi} \cup \boldsymbol{\pi}', s) \mid \rho \models \psi\}) + Pr_{\mathcal{G}(\boldsymbol{\pi} \cup \boldsymbol{\pi}', s)}(\{\rho \in \mathcal{G}(\boldsymbol{\pi} \cup \boldsymbol{\pi}', s) \mid \rho \models \neg \psi\}) = 1$. Therefore $Pr_{\mathcal{G}(\boldsymbol{\pi} \cup \boldsymbol{\pi}', s)}(\{\rho \in \mathcal{G}(\boldsymbol{\pi} \cup \boldsymbol{\pi}', s) \mid \rho \models \psi\}) \bowtie \alpha$ iff $Pr_{\mathcal{G}(\boldsymbol{\pi} \cup \boldsymbol{\pi}', s)}(\{\rho \in \mathcal{G}(\boldsymbol{\pi} \cup \boldsymbol{\pi}', s) \mid \rho \models \neg \psi\}) \widetilde{\bowtie} 1 - \alpha$, for all $\alpha \in [0, 1]$ and $\bowtie \in \{<, >, \leq, \geq\}$. □

Therefore, the path quantifier R (release) can be expressed by the existing PATL syntax, in the way that $\langle\!\langle A \rangle\!\rangle^{\bowtie \alpha} \phi_1 \mathtt{R}^{\leq k} \phi_2 \equiv \langle\!\langle A \rangle\!\rangle^{\widetilde{\bowtie} 1 - \alpha} (\neg \phi_1) \mathtt{U}^{\leq k} (\neg \phi_2)$, where both $\neg \phi_1$ and $\neg \phi_2$ are state formulas.

**On Model Checking of PATL**

In this section we briefly survey the results in the literature related to PATL model checking. Given a PATL formula in the form of $\langle\!\langle A \rangle\!\rangle^{\bowtie \alpha} \psi(\phi_1, \ldots, \phi_n)$, regarding $[\![\phi_1]\!], \ldots, [\![\phi_n]\!]$ as the sets of states satisfying state formulas $\phi_1, \ldots, \phi_n$, a standard way to solve this problem is to determine the maximal or minimal probability that the players in $A$ can enforce the LTL formula $\psi(\phi_1, \ldots, \phi_n)$. In the following we write $\psi$ for $\psi(\phi_1, \ldots, \phi_n)$ without further confusions.

LTL properties are special cases of $\omega$-regular winning objectives [Tho91] in two-player concurrent (zero-sum) games [dAM04,CdAH06]. In such games one may group a set of players $A \subseteq \Sigma$ into a single protagonist and $\overline{A}$ into a single antagonist. Given an $\omega$-regular winning objective $\xi$ and starting from a state $s \in S$, the protagonist plays with a strategy trying to maximize the probability for a play to satisfy $\xi$ while the antagonist tries to minimize the probability. In such a game there always exists a unique value in $[0, 1]$, on which both players have strategies to guarantee (or infinitely approach) their best performances, regardless of the strategies played by their opponents. Such a supremum value (or infinum value, as for the antagonist) is called the *value of the game* [Mar98,dAM04]. In a probabilistic multi-player game, we let a group of players $A \subseteq \Sigma$ be a single player, and $\overline{A}$ be the other, and the supremal probability for $A$ to enforce such an LTL formula $\psi$ starting from a given state $s \in S$ can be uniquely determined. This value is defined as

$$\langle A \rangle \psi(s) = \bigsqcup_{\pi \in \Pi(A)} \bigsqcap_{\pi' \in \Pi(\overline{A})} Pr_{\mathcal{G}(\pi \cup \pi', s)}(\{\rho \in \mathcal{G}(\pi \cup \pi', s) \mid \rho \models \psi\})$$

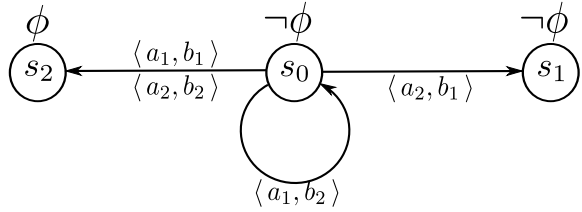

**Fig. 1.** An example showing that player I can guarantee to satisfy $\Diamond\phi$ with probability $\alpha$ for all $0 \leq \alpha < 1$, but he cannot ensure that property with probability 1.

*Example 1.* Fig. 1 gives a PGS with two players $\{\text{I}, \text{II}\}$, initial state $s_0$, $\text{Act}_\text{I} = \{a_1, a_2\}$ and $\text{Act}_\text{II} = \{b_1, b_2\}$. Note that this PGS is deterministic, i.e, no probabilities in its transitions. We assume that the only available transitions from $s_1$ and $s_2$ are self-loops, and the other transition relations are as depicted in the graph. Suppose player I wants to maximize the probability to enforce the property $\Diamond\phi$, and player II aims to minimize it.

Since the strategies applied on $s_1$ and $s_2$ do not matter, we focus on the choices of actions from both players on $s_0$. We first focus on memoryless strategies, and let player I's strategy $\pi_1$ gives $\pi_1(\gamma)(a_1) = p$ and $\pi_1(\gamma)(a_2) = 1 - p$ for all $\gamma \in S^+$. Similarly we let II assign probability $q$ to $b_1$ and $1 - q$ to $b_2$ all the time. This produces an infinite tree, on which we write $x_{s_0}(\text{I})$ for the actual probability I can achieve $\Diamond\phi$ from $s_0$, given the above memoryless strategies. (Note that $x_{s_1}(\text{I}) = 0$ and $x_{s_2}(\text{I}) = 1$ in all cases.) This establishes an equation which further derives $x_{s_0}(\text{I}) = \frac{(1-p)+(2p-1)q}{(1-p)+pq}$. A simple analysis shows that when $p$ approaches 1, the minimal value of $x_{s_0}(\text{I})$ approaches 1 as well, for all choices of $q$. That is, there exists a strategy for player I to enforce $\Diamond\phi$ with probability $1 - \varepsilon$ for all $\varepsilon > 0$. However, if player I chooses $p = 1$, player II may set $q = 0$ so that a play will be trapped in $s_0$ for ever that yields $x_{s_0}(\text{I}) = 0$. The result of [dAM04] shows that in this case player I cannot do better even with general

(history dependent) strategies. In fact there are no strategies for player I to enforce $\Diamond\phi$ with probability 1. □

Indeed, $\langle A\rangle\psi(s)$ can be *almost* the best, i.e., we have $\mathcal{G}, s \models \langle\!\langle A\rangle\!\rangle^{\geq \langle A\rangle\psi(s)-\varepsilon}\psi$ for all $\varepsilon > 0$ [dAHK98]. Nevertheless, the quantitative version of determinacy [Mar98] ensures that for all LTL formulas $\psi$ and $s \in S$, we have

$$\langle A\rangle\psi(s) + \langle\overline{A}\rangle\neg\psi(s) = 1$$

The PATL model checking problems can be solved by calculating the values $\langle A\rangle\psi_s(s)$ for each state $s$, where each local objective $\psi_s$ related to $s$ might be distinct. The algorithms of [dAM04] define monotonic functions of type $(S \to [0,1]) \to (S \to [0,1])$ to arbitrarily approach a vector $\{\langle A\rangle\psi_s(s)\}_{s\in S}$ in a game structure with finite state space $S$ with respect to an $\omega$-regular winning objective $\psi$. Within each step one has to go through $\mathcal{O}(|S|)$ matrix games, and each iteration produces a unique fixed point. The algorithms on safety and reachability objectives are special cases of solving stochastic games [RF91]. More complex properties can be expressed as nested fixed points [dAM04]. Therefore, the upper bound complexities become exponential to the size of the winning objectives translated from LTL formulas. More recently, alternative algorithms proposed in [CdAH06] prove that for quantitative games with $\omega$-regular winning objectives expressed as parity conditions, whether the values of a game is within $[r-\epsilon, r+\epsilon]$ can be decided in $NP \cap coNP$ for all rational $r \in [0,1]$ and $\epsilon > 0$, which improves the theoretical upper bound for estimating the optimal values.

**Optimal Strategies**

It has been shown in [dAM04] that for *safety* games there always exist optimal strategies for the protagonists, however for *reachability* games it is not always the case. As shown in example 1, player I has no optimal strategy to enforce $\Diamond\phi$ with probability 1 on $s_0$ even though $\langle\texttt{I}\rangle\phi(s_0) = 1$. Based on similar proof strategies applied in [dAM04], we examine the existence of optimal strategies on winning objectives expressed as path formulas of PATL on a state.

**Lemma 2.** *Let $s$ be a state, $\psi$ be a path formula, and $A$ the set of protagonists.*

1. *If $\psi$ is of the form $\bigcirc\phi$, $\phi_1\mathtt{U}^{\leq k}\phi_2$, $\phi_1\mathtt{R}^{\leq k}\phi_2$, or $\phi_1\mathtt{R}\phi_2$ with $k \in \mathbb{N}$, there always exists a joint optimal strategy for $A$ that enforces $\psi$ on $s$ with probability at least $\langle A\rangle\psi(s)$.*
2. *If $\psi$ is of the form $\phi_1\mathtt{U}\phi_2$, there always exists a joint $\epsilon$-optimal strategy for $A$ that enforces $\psi$ on $s$ with probability at least $\langle A\rangle\psi(s) - \epsilon$, for all $\epsilon > 0$.*

For the prove of Lemma 2 we rely on the representation of a solution for a winning objective in quantitative game $\mu$-calculus [dAM04]. For the sake of readability we leave the whole proof in the appendix.

The next result proves the existence of a joint $A$ strategy to enforce an PATL path formula with probability greater than $\alpha$ if there exists a joint strategy to enforce that formula with probability greater than $\alpha$ against an optimal $\overline{A}$ strategy.

**Lemma 3.** *Let $\psi$ be a PATL path formula and $\pi'$ be a joint optimal strategy for the antagonists $\overline{A}$ on state $s$, if there exists a joint strategy $\pi$ for the protagonists $A$ such that $Pr_{\mathcal{G}(\pi \cup \pi', s)}(\{\rho \in \mathcal{G}(\pi \cup \pi', s) \mid \rho \models \psi\}) > \alpha$, then $\mathcal{G}, s \models \langle\!\langle A \rangle\!\rangle^{>\alpha} \psi$.*

*Proof.* Since $\pi'$ is the optimal strategy for the antagonists, we have for all joint strategies $\pi''$, $Pr_{\mathcal{G}(\pi'' \cup \pi', s)}(\{\rho \in \mathcal{G}(\pi'' \cup \pi', s) \mid \rho \models \psi\}) \leq \langle A \rangle \psi(s)$, then we have $\langle A \rangle \psi(s) > \alpha$. If there exists an optimal joint strategy for $A$ then we have $s \models \langle\!\langle A \rangle\!\rangle^{\geq \langle A \rangle \psi(s)} \psi$, which implies $s \models \langle\!\langle A \rangle\!\rangle^{>\alpha} \psi$. Otherwise by Lemma 2 there exists an $\epsilon$-optimal joint strategy for $A$ with small $\epsilon > 0$ to enforce $\psi$ with probability at least $\langle A \rangle \psi(s) - \epsilon > \alpha$. This also gives us $s \models \langle\!\langle A \rangle\!\rangle^{>\alpha} \psi$. □

This result does not hold if we replace the operator ">" by "≥" for unbounded until U. This is because if there does not exist a joint optimal strategy for $A$ to enforce $\phi_1 \mathtt{U} \phi_2$ with probability $\geq \alpha$, we have no space to insert a tiny $\epsilon > 0$ as we did in the above proof. For the fragment of path formulas without unbounded until, we extend the results for $\geq$, by the fact that optimal joint strategies for $A$ always exist for these path modalities, as shown by Lemma 2.

**Lemma 4.** *For path formulas $\psi$ in the form of $\bigcirc \phi$ or $\phi_1 \mathtt{U}^{\leq k} \phi_2$ and optimal strategies $\pi'$ for the antagonists $\overline{A}$ on state $s$, if there exists a joint strategy $\pi$ for the protagonists $A$ such that $Pr_{\mathcal{G}(\pi \cup \pi', s)}(\{\rho \in \mathcal{G}(\pi \cup \pi', s) \mid \rho \models \psi\}) \bowtie \alpha$, then $\mathcal{G}, s \models \langle\!\langle A \rangle\!\rangle^{\bowtie \alpha} \psi$, where $k \in \mathbb{N}$ and $\bowtie \in \{>, \geq\}$.*

*Proof.* Since there exists joint strategies for $A$ against $\overline{A}$'s optimal strategies, we have $\langle \overline{A} \rangle \neg \psi(s) \widetilde{\bowtie} 1 - \alpha$, therefore $\langle A \rangle \psi(s) \bowtie \alpha$ by determinacy. By Lemma 2 there always exist optimal strategies for $A$ to enforce $\psi$ with probability $\bowtie \alpha$ if $\psi$ is in the form of $\bigcirc \phi$ or $\phi_1 \mathtt{U}^{\leq k} \phi_2$. □

### *A*-PATL

We define a sublogic of PATL by focusing on a particular set of players. Similar to the approach of [AHKV98], we only allow negations to appear on the level of propositions. Let $A \subseteq \Sigma$, an *A*-PATL formula $\phi$ is a state formula defined as follows:

$$\phi := p \mid \neg p \mid \phi_1 \wedge \phi_2 \mid \phi_1 \vee \phi_2 \mid \langle\!\langle A' \rangle\!\rangle^{\bowtie \alpha} \bigcirc \phi \mid \langle\!\langle A' \rangle\!\rangle^{\bowtie \alpha} \phi_1 \mathtt{U}^{\leq k} \phi_2 \mid \langle\!\langle A' \rangle\!\rangle^{>\alpha} \phi_1 \mathtt{U} \phi_2$$

where $k \in \mathbb{N}$, $\bowtie \in \{>, \geq\}$ and $A' \subseteq A$. Write $\mathbb{L}_A$ for the set of *A*-PATL formulas. An *A*-PATL formula describes a property that players in $A$ are able to ensure with a minimal expectation by their joint strategies. Note that we only allow '$> \alpha$' in the construction of unbounded until.

## 5 Probabilistic Alternating Simulation Relations

We define probabilistic versions of alternating simulation [AHKV98]. An alternating simulation is a two-step simulation. For a sketch, suppose state $s$ is simulated by state $t$. In the first step the protagonists choose their actions on $t$ to simulate the behaviour of the protagonists on $s$, and in the second step the antagonists choose actions on $s$ to respond the behaviour of the antagonists on $t$. This somehow results in a simulation-like relation, so that for a certain property the protagonists can enforce on $s$, they can also enforce it on $t$. To this end we split $\Sigma$ into two groups of players — one group of protagonist and the other group of antagonist. Subsequently, we consider only the two-player case in a probabilistic game structure — player I for the protagonist and player II for the antagonist, since what we can achieve in the two-player case naturally extends to a result in systems with two complementary sets of players, i.e., $A \cup \overline{A} = \Sigma$. For readability we also write the transition functions as $\delta(s, a_1, a_2)$ and $\overline{\delta}(s, \pi_1, \pi_2)$ for $\delta(s, \langle a_1, a_2 \rangle)$ and $\overline{\delta}(s, \langle \pi_1, \pi_2 \rangle)$, respectively.

Let $S, T$ be two sets and $\mathcal{R} \subseteq S \times T$ be a relation, then $\overline{\mathcal{R}} \subseteq \mathcal{D}(S) \times \mathcal{D}(T)$ is defined by $\Delta \overline{\mathcal{R}} \Theta$ if there exists a weight function $w : S \times T \to [0, 1]$ satisfying

- $\sum_{t \in T} w(s, t) = \Delta(s)$ for all $s \in S$,
- $\sum_{s \in S} w(s, t) = \Theta(t)$ for all $t \in T$,
- $s \mathcal{R} t$ for all $s \in S$ and $t \in T$ with $w(s, t) > 0$.

Note in this definition, it is equivalent to have $\sum_{t \in \lceil \Theta \rceil} w(s, t) = \Delta(s)$ for all $s \in S$, and $\sum_{s \in \lceil \Delta \rceil} w(s, t) = \Theta(t)$ for all $t \in T$. Since $w$ can only assign non-zero values to the states in the support of $\Delta$ or $\Theta$. If $w(s, t) > 0$ for some $s \notin \lceil \Delta \rceil$ and $t \in T$, then we would have $\sum_{t \in T} w(s, t) > 0 = \Delta(s)$, which is a contradiction. The followings are several properties of lifted relations.

**Lemma 5.** *(inverse) Let $\mathcal{R}^{-1} \subseteq T \times S$ be the inverse of $R \subseteq S \times T$, then for all $\Delta \in \mathcal{D}(S)$ and $\Theta \in \mathcal{D}(T)$, $\Delta \overline{\mathcal{R}} \Theta$ iff $\Theta \overline{\mathcal{R}^{-1}} \Delta$.*

*Proof.* By taking the inverse of the weight function. □

**Lemma 6.** *Let $\Delta \in \mathcal{D}(S)$, $\Delta' \in \mathcal{D}(S')$, and $\mathcal{R}$ a relation on $S$. If $\Delta \overline{\mathcal{R}} \Delta'$, then*

1. *If there exist $\Delta_1, \Delta_2, \cdots \in \mathcal{D}(S)$ and an index set $\{p_i\}_I$ satisfying $\sum_{i \in I} p_i = 1$ and $\Delta = \sum_{i \in I} p_i \cdot \Delta_i$, then there exist $\Delta'_1, \Delta'_2 \cdots \in \mathcal{D}(S')$ such that $\Delta' = \sum_{i \in I} p_i \cdot \Delta'_i$, and $\Delta_i \overline{\mathcal{R}} \Delta'_i$ for all $i \in I$.*
2. *If there exist $\Delta'_1, \Delta'_2, \cdots \in \mathcal{D}(S')$ and an index set $\{p_i\}_I$ satisfying $\sum_{i \in I} p_i = 1$ and $\Delta' = \sum_{i \in I} p_i \cdot \Delta'_i$, then there exist $\Delta_1, \Delta_2 \cdots \in \mathcal{D}(S)$ such that $\Delta = \sum_{i \in I} p_i \cdot \Delta_i$, and $\Delta_i \overline{\mathcal{R}} \Delta'_i$ for all $i \in I$.*

*Proof.* We prove the second part, and the first part is similar. Let $\Delta' = \sum_{i \in I} p_i \cdot \Delta'_i$, then define $\Delta_i$ for each $i \in I$ by $\Delta_i(s) = \sum_{s' \in S'} w(s, s') \cdot \frac{\Delta'_i(s')}{\Delta'(s')}$ for all $s \in S$. Now we can check that $\sum_{i \in I} p_i \cdot \Delta_i(s) = \Delta(s)$ for all $s$, i.e., $\Delta = \sum_{i \in I} p_i \cdot \Delta_i$.

To show that $\Delta_i \overline{\mathcal{R}} \Delta'_i$, we define a weight function $w_i : S \times S' \to [0, 1]$ by for all $s \in S$ and $s' \in S'$, $w_i(s, s') = w(s, s') \cdot \frac{\Delta'_i(s')}{\Delta'(s')}$. Consider the following conditions.

1. $w_i(s, s') > 0$ implies $w(s, s') > 0$, therefore $s \mathcal{R} s'$.
2. For all $s \in S$, we have $\sum_{s' \in S'} w_i(s, s') = \sum_{s' \in S'} w(s, s') \cdot \frac{\Delta'_i(s')}{\Delta'(s')} = \Delta_i(s)$,

3. For all $s' \in S'$, we have
   $\sum_{s \in S} w_i(s, s') = \sum_{s \in S} w(s, s') \cdot \frac{\Delta'_i(s')}{\Delta'(s')} = \frac{\Delta'_i(s')}{\Delta'(s')} \cdot \sum_{s \in S} w(s, s') = \frac{\Delta'_i(s')}{\Delta'(s')} \cdot \Delta'(s') = \Delta'_i(s')$.
   □

**Lemma 7.** *Let $\mathcal{R}$ be a relation on $S$ and $\{p_i\}_{i \in I}$ be an index set satisfying $\sum_{i \in I} p_i = 1$ and $\Delta_i \overline{\mathcal{R}} \Delta'_i$ for distributions $\Delta_i, \Delta'_i \in \mathcal{D}(S)$ for all $i$, then $\sum_{i \in I} p_i \cdot \Delta_i \overline{\mathcal{R}} \sum_{i \in I} p_i \cdot \Delta'_i$.*

*Proof.* W.l.o.g., let $\Delta_i \in \mathcal{D}(S)$ and $\Delta'_i \in \mathcal{D}(S')$ for all $i$, and let $w_i$ be the weight function for $\Delta_i \overline{\mathcal{R}} \Delta'_i$. Define a new weight function $w : S \times S' \to [0, 1]$, by $w(s, s') = \sum_{i \in I} p_i \cdot w_i(s, s')$.

- $w(s, s') > 0$, then $\sum_{i \in I} p_i \cdot w_i(s, s') > 0$, i.e, there exists some $i \in I$ such that $w_i(s, s') > 0$, which gives $s\mathcal{R}s'$.
- For all $s \in S$, $\sum_{s' \in S'} w(s, s')$
  $= \sum_{s' \in S'} \sum_{i \in I} p_i \cdot w_i(s, s')$
  $= \sum_{i \in I} p_i \cdot \sum_{s' \in S'} w_i(s, s')$
  $= \sum_{i \in I} p_i \cdot \Delta_i(s)$.
- To show that for all $s' \in S'$, $\sum_{s \in S} w(s, s') = \sum_{i \in I} p_i \cdot \Delta'_i(s')$ is similar.

This gives $\sum_{i \in I} p_i \cdot \Delta_i \overline{\mathcal{R}} \sum_{i \in I} p_i \cdot \Delta'_i$.
□

Based on the notion of lifting, we define the probabilistic alternating simulation relation for player I that extends the alternating simulation relation of [AHKV98]. The definition for player II can be made in a similar way.

**Definition 1.** *Consider $\mathcal{G}, \mathcal{G}'$ as two probabilistic game structures. A probabilistic alternating I-simulation $\sqsubseteq \subseteq S \times S'$ is a relation satisfying if $s \sqsubseteq s'$, then*

- $\mathcal{L}(s) = \mathcal{L}'(s')$,
- *for all $\pi_1 \in \Pi_I^{\mathcal{G},1}$, there exists $\pi'_1 \in \Pi_I^{\mathcal{G}',1}$, such that for all $\pi'_2 \in \Pi_{II}^{\mathcal{G}',1}$, there exists $\pi_2 \in \Pi_{II}^{\mathcal{G},1}$, such that $\overline{\delta}(s, \pi_1, \pi_2) \overline{\sqsubseteq} \, \overline{\delta'}(s', \pi'_1, \pi'_2)$.*

Let $\mathcal{R} \subseteq S \times S'$ and $\mathcal{R}' \subseteq S' \times S''$ be two relations, then $\mathcal{R} \cdot \mathcal{R}'$ is a relation on $S \times S''$ defined by $s(\mathcal{R} \cdot \mathcal{R}') \, s''$ if there exists $s' \in S'$ such that $s\mathcal{R}s'$ and $s'\mathcal{R}'s''$.

**Lemma 8.** *(Transitivity of alternating simulation) Consider $\mathcal{G}, \mathcal{G}'$ and $\mathcal{G}''$ be three probabilistic game structures. If $\sqsubseteq \subseteq S \times S'$ and $\sqsubseteq' \subseteq S' \times S''$ are probabilistic alternating I-simulations, then $\sqsubseteq \cdot \sqsubseteq'$ is a probabilistic alternating I-simulation on $S \times S''$.*

*Proof.* (sketch) Let $s \sqsubseteq \cdot \sqsubseteq' s''$, then by definition there exists $s' \in S'$ such that $s \sqsubseteq s'$ and $s' \sqsubseteq' s''$. Therefore $\mathcal{L}(s) = \mathcal{L}(s') = \mathcal{L}(s'')$. Let $\pi_1 \in \Pi_I^{\mathcal{G},1}$, then by definition there exists $\pi'_1 \in \Pi_I^{\mathcal{G}',1}$ such that for all $\pi'_2 \in \Pi_{II}^{\mathcal{G}',1}$ there exists $\pi_2 \in \Pi_{II}^{\mathcal{G},1}$ such that $\overline{\delta}(s, \langle \pi_1, \pi_2 \rangle) \overline{\sqsubseteq} \overline{\delta'}(s', \langle \pi'_1, \pi'_2 \rangle)$. By $s' \sqsubseteq' s''$, there exists $\pi''_1 \in \Pi_I^{\mathcal{G}'',1}$ such that for all $\pi''_3 \in \Pi_{II}^{\mathcal{G}'',1}$, there exists $\pi'_3 \in \Pi_{II}^{\mathcal{G}',1}$ such that $\overline{\delta'}(s', \langle \pi'_1, \pi'_3 \rangle) \overline{\sqsubseteq'} \, \overline{\delta''}(s'', \langle \pi''_1, \pi''_3 \rangle)$. Then from above there also exists

$\pi_3 \in \Pi_\mathtt{I}^{\mathcal{G},1}$ such that $\overline{\delta}(s, \langle \pi_1, \pi_3 \rangle) \overline{\sqsubseteq \delta'}(s', \langle \pi_1', \pi_3' \rangle)$. Write $\Delta = \overline{\delta}(s, \langle \pi_1, \pi_3 \rangle)$, $\Delta' = \overline{\delta'}(s', \langle \pi_1', \pi_3' \rangle)$ and $\Delta'' = \overline{\delta''}(s'', \langle \pi_1'', \pi_3'' \rangle)$. we need to show that $\Delta \overline{\sqsubseteq \cdot \sqsubseteq'} \Delta''$.

Let $w_1$ be a weight function for $\Delta \overline{\sqsubseteq} \Delta'$ and $w_2$ a weight function for $\Delta' \overline{\sqsubseteq'} \Delta''$, define a new weight function $w : S \times S'' \to [0,1]$, by $w(s, s'') = \sum_{s' \in S'} \frac{w_1(s,s') \cdot w_2(s',s'')}{\Delta'(s')}$. Let $s \in S$ and $s'' \in S''$.

- If $w(s, s'') > 0$ then exists $s' \in \lceil \Delta' \rceil$ such that $w_1(s, s') > 0$ and $w_2(s', s'') > 0$, which implies $s \sqsubseteq s'$ and $s' \sqsubseteq' s''$. Therefore, $s \sqsubseteq \cdot \sqsubseteq' s''$.
- $\sum_{s \in S} w(s, s'')$
  $= \sum_{s \in S} \sum_{s' \in S'} \frac{w_1(s,s') \cdot w_2(s',s'')}{\Delta'(s')}$
  $= \sum_{s' \in S'} \frac{w_2(s',s'')}{\Delta'(s')} \cdot \sum_{s \in S} w_1(s, s')$
  $= \sum_{s' \in S'} \frac{w_2(s',s'')}{\Delta'(s')} \cdot \Delta'(s')$
  $= \sum_{s' \in S'} w_2(s', s'')$
  $= \Delta''(s'')$
- Showing $\sum_{s'' \in S''} w(s, s'') = \Delta(s)$ is similar.

□

Lemma 8 can also be derived from the transitivity of probabilistic alternating forward simulation (Corollary 1) and the fact that every probabilistic alternating simulation is also a probabilistic alternating forward simulation (Lemma 11).

Based on the probabilistic forward simulation of Segala [Seg95a], and the alternating simulation of Alur et al. [AHKV98], we propose the notion of probabilistic alternating forward simulation. A forward simulation relates a state to a distribution of states, which requires a different way of lifting. Let $\mathcal{R} \subseteq S \times \mathcal{D}(S)$ be a relation, write $\overline{\mathcal{R}}$ for the smallest relation satisfying $\Delta \overline{\mathcal{R}} \Theta$ if there exists an index set $\{p_i\}_{i \in I}$ satisfying $\Sigma_{i \in I} p_i = 1$, such that $\Delta = \Sigma_{i \in I} p_i \cdot \overline{s_i}$, $\Theta = \Sigma_{i \in I} p_i \cdot \Theta_i$ and $s_i \mathcal{R} \Theta_i$ for all $i$. We call $\overline{\mathcal{R}}$ the *forward lifting* of $\mathcal{R}$. Forward lifting has the following similar properties as the previous lifting.

**Lemma 9.** *Let $\mathcal{R}$ be a relation on $S \times \mathcal{D}(S)$ and $\{p_i\}_{i \in I}$ be an index set satisfying $\sum_{i \in I} p_i = 1$ and $\Delta_i \overline{\mathcal{R}} \Delta_i'$ for distributions $\Delta_i, \Delta_i' \in \mathcal{D}(S)$ for all $i$, then $\sum_{i \in I} p_i \cdot \Delta_i \overline{\mathcal{R}} \sum_{i \in I} p_i \cdot \Delta_i'$, where $\overline{\mathcal{R}}$ is the forward lifting of $\mathcal{R}$.*

**Lemma 10.** *Let $\Delta \in \mathcal{D}(S)$, $\Delta' \in \mathcal{D}(S')$, and $\mathcal{R}$ a relation on $S$. If $\Delta \overline{\mathcal{R}} \Delta'$, and there exist $\Delta_1, \Delta_2, \cdots \in \mathcal{D}(S)$ and an index set $\{p_i\}_I$ satisfying $\sum_{i \in I} p_i = 1$ and $\Delta = \sum_{i \in I} p_i \cdot \Delta_i$, then there exist $\Delta_1', \Delta_2' \cdots \in \mathcal{D}(S')$ such that $\Delta' = \sum_{i \in I} p_i \cdot \Delta_i'$, and $\Delta_i \overline{\mathcal{R}} \Delta_i'$ for all $i \in I$, where $\overline{\mathcal{R}}$ is the forward lifting of $\mathcal{R}$.*

Now we define the probabilistic alternating forward simulation relation for player I, and the definition for player II can be made in a similar way.

**Definition 2.** *Consider two probabilistic game structures $\mathcal{G} = \langle S, s_0, \mathcal{L}, \mathtt{Act}, \delta \rangle$ and $\mathcal{G}' = \langle S', s_0', \mathcal{L}', \mathtt{Act}', \delta' \rangle$. A probabilistic alternating forward I-simulation $\sqsubseteq_\mathsf{f} \subseteq S \times \mathcal{D}(S')$ is a relation satisfying if $s \sqsubseteq_\mathsf{f} \Delta'$, then*

- *$\mathcal{L}(s) = \mathcal{L}'(s')$ for all $s' \in \lceil \Delta' \rceil$,*

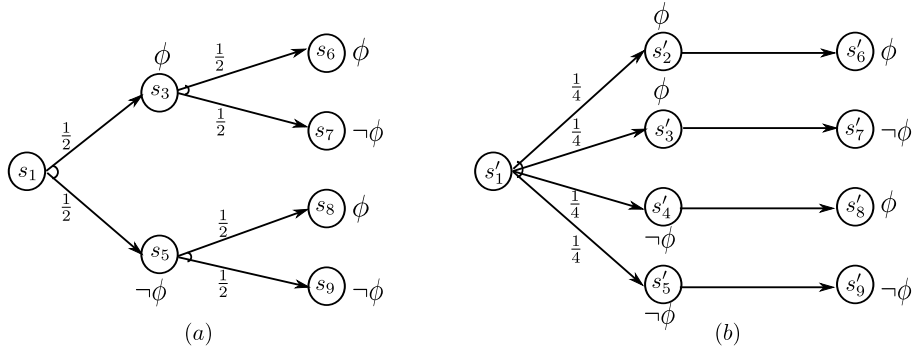

**Fig. 2.** An example showing that probabilistic alternating forward simulation is strictly weaker than probabilistic alternating simulation.

- for all $\pi_1 \in \Pi_{\text{I}}^{\mathcal{G},1}$, there exists $\pi_1' \in \Pi_{\text{I}}^{\mathcal{G}',1}$, such that for all $\pi_2' \in \Pi_{\text{II}}^{\mathcal{G}',1}$, there exists $\pi_2 \in \Pi_{\text{II}}^{\mathcal{G},1}$, such that $\overline{\delta}(s, \pi_1, \pi_2) \overline{\sqsubseteq}_{\text{f}} \overline{\delta'}(\Delta', \pi_1', \pi_2')$.

**Lemma 11.** $s \sqsubseteq t$ implies $s \sqsubseteq_{\text{f}} \overline{t}$.

This lemma says that every probabilistic alternating simulation is a probabilistic forward simulation with a point distribution on the right hand side of the relation. The other way does not hold, i.e., probabilistic alternating forward simulation relates strictly more game structures than probabilistic alternating simulation. In Fig. 2, we assume $\text{Act}_{\text{I}}$ and $\text{Act}_{\text{II}}$ are both singleton sets. One may find that there are no states in the set $\{s_2', s_3', s_4', s_5'\}$ in Fig. 2(b) that can simulate states $s_3$ and $s_5$ in Fig. 2(a). Therefore, we cannot establish a probabilistic alternating simulation from $s_1$ to $s_1'$. However, $s_1$ is related to $s_1'$ by probabilistic alternating forward simulation, since $s_3$ ($s_5$) can be related to a uniform distribution over $s_2'$ and $s_3'$ ($s_4'$ and $s_5'$).

Before proceeding to the next step we introduce the following several auxiliary lemmas.

**Lemma 12.** Given $\{p_i\}_{i \in I}$ with $\sum_{i \in I} p_i = 1$, $s \in S$, $\{\pi_i\}_{i \in I} \subseteq \Pi_{\text{I}}^{\mathcal{G},1}$ and $\pi \in \Pi_{\text{II}}^{\mathcal{G},1}$, we have $\overline{\delta}(s, \sum_{i \in I} p_i \cdot \pi_i, \pi) = \sum_{i \in I} p_i \cdot \overline{\delta}(s, \pi_i, \pi)$.

*Proof.* Let $t \in S$, then we have

$$\begin{aligned}
\overline{\delta}(s, \sum_{i \in I} p_i \cdot \pi_i, \pi)(t) &= \delta(s, \sum_{i \in I} p_i \cdot \pi_i(s), \pi(s))(t) \\
&= \sum_{a_1 \in \text{Act}_1} \sum_{a_2 \in \text{Act}_2} \sum_{i \in I} p_i \cdot \pi_i(a_1) \cdot \pi(a_2) \cdot \delta(s, a_1, a_2)(t) \\
&= \sum_{i \in I} p_i \cdot [\sum_{a_1 \in \text{Act}_1} \sum_{a_2 \in \text{Act}_2} \pi_i(a_1) \cdot \pi(a_2) \cdot \delta(s, a_1, a_2)(t)] \\
&= \sum_{i \in I} p_i \cdot \delta(s, \pi_i(s), \pi(s))(t) \\
&= \sum_{i \in I} p_i \cdot \overline{\delta}(s, \pi_i, \pi)(t)
\end{aligned}$$

□

**Lemma 13.** *Let $\mathcal{G}$ and $\mathcal{G}'$ be two game structures, $\{\Delta_i\}_{i \in I}$ be a set of distributions, $\{\pi_i\}_{i \in I}$ a set of level 1 $\mathtt{I}$-strategies, $\pi \in \Pi_{\mathtt{II}}^{\mathcal{G}',1}$ and $\{p_i\}_{i \in I}$ satisfies $\sum_{i \in I} p_i = 1$, we have $\overline{\delta}(\sum_{i \in I} p_i \cdot \Delta_i, \pi', \pi) = \sum_{i \in I} p_i \cdot \overline{\delta}(\Delta_i, \pi_i, \pi)$, where $\pi'$ is a level 1 $\mathtt{I}$ strategy defined by*

$$\pi'(s) = \sum_{i \in I} \frac{p_i \cdot \Delta_i(s) \cdot \pi_i(s)}{\sum_{i \in I} p_i \cdot \Delta_i(s)}$$

*Proof.* Write $LHS = \overline{\delta}(\sum_{i \in I} p_i \cdot \Delta_i, \pi', \pi)$ and $RHS = \sum_{i \in I} p_i \cdot \overline{\delta}(\Delta_i, \pi_i, \pi)$, we need to show for all $t \in S$, $LHS(t) = RHS(t)$. Write $\Delta$ for the distribution $\sum_{i \in I} p_i \cdot \Delta_i$. Let $t \in S$, then by definition

$$\begin{aligned}
LHS(t) &= \sum_{s \in \lceil \Delta \rceil} \sum_{i \in I} p_i \cdot \Delta_i(s) \overline{\delta}(s, \pi', \pi)(t) \\
&= \sum_{s \in \lceil \Delta \rceil} \Delta(s) \cdot \overline{\delta}(s, \pi', \pi)(t) \\
&= \sum_{s \in \lceil \Delta \rceil} \Delta(s) \cdot \delta(s, \pi'(s), \pi(s))(t) \\
&= \sum_{s \in \lceil \Delta \rceil} \Delta(s) \cdot \delta(s, \sum_{j \in I} \frac{p_j \cdot \Delta_j(s) \pi_j(s)}{\Delta(s)}, \pi(s))(t) \\
&= \sum_{s \in \lceil \Delta \rceil} \Delta(s) \cdot \sum_{j \in I} \frac{p_j \cdot \Delta_j(s)}{\Delta(s)} \delta(s, \pi_j(s), \pi(s))(t) & \text{By Lemma 12} \\
&= \sum_{s \in \lceil \Delta \rceil} \cdot \sum_{j \in I} p_j \cdot \Delta_j(s) \cdot \delta(s, \pi_j(s), \pi(s))(t) \\
&= \sum_{j \in I} p_j \cdot \sum_{s \in \lceil \Delta \rceil} \cdot \Delta_j(s) \cdot \delta(s, \pi_j(s), \pi(s))(t) \\
&= \sum_{j \in I} p_j \cdot \sum_{s \in \lceil \Delta_j \rceil} \cdot \Delta_j(s) \cdot \delta(s, \pi_j(s), \pi(s))(t) \\
&= \sum_{j \in I} p_j \cdot \overline{\delta}(\Delta_i, \pi_i, \pi)(t) & \text{By definition} \\
&= RHS(t)
\end{aligned}$$

$\square$

The next result shows that the definition of forward simulation also works on the lifted relation.

**Lemma 14.** *If $\Delta \overline{\sqsubseteq_{\mathsf{f}}} \Theta$, then for all $\pi_1 \in \Pi_{\mathtt{I}}^{\mathcal{G},1}$, there exists $\pi_2 \in \Pi_{\mathtt{I}}^{\mathcal{G}',1}$, such that for all $\pi_2' \in \Pi_{\mathtt{II}}^{\mathcal{G}',1}$, there exists $\pi_1' \in \Pi_{\mathtt{II}}^{\mathcal{G},1}$, such that $\overline{\delta}(\Delta, \pi_1, \pi_1') \overline{\sqsubseteq_{\mathsf{f}}} \overline{\delta}(\Theta, \pi_2, \pi_2')$.*

*Proof.* By definition there exists a set $\{p_i\}_{i \in I}$ such that $\Delta = \sum_{i \in I} p_i \cdot \overline{s}_i$, $\Theta = \sum_{i \in I} p_i \cdot \Theta_i$ and $s_i \sqsubseteq_{\mathsf{f}} \Theta_i$. Let $\pi_1 \in \Pi_{\mathtt{I}}^{\mathcal{G},1}$ be a (mixed) $\mathtt{I}$-strategy. Then by definition, for all $i \in I$ there exists $\pi_i \in \Pi_{\mathtt{I}}^{\mathcal{G}',1}$ such that for all $\pi_i' \in \Pi_{\mathtt{II}}^{\mathcal{G}',1}$, there exists $\pi_i'' \in \Pi_{\mathtt{II}}^{\mathcal{G},1}$ such that $\overline{\delta}(s_i, \pi_1, \pi_i'') \overline{\sqsubseteq_{\mathsf{f}}} \overline{\delta}(s_i, \pi_i, \pi_i')$. Now we take $\pi_2$ defined by $\pi_2(s) = \sum_{i \in I} p_i \frac{\Theta_i(s)}{\Theta(s)} \pi_i(s)$ for all $s$ be the required (level 1) $\mathtt{I}$-strategy.

Let $\pi_2' \in \Pi_{\mathtt{II}}^{\mathcal{G}',1}$, we prove as follows the existence of another $\mathtt{II}$-strategy $\pi_1' \in \Pi_{\mathtt{II}}^{\mathcal{G},1}$ that satisfies $\overline{\delta}(\Delta, \pi_1, \pi_1') \overline{\sqsubseteq_{\mathsf{f}}} \overline{\delta}(\Theta, \pi_2, \pi_2')$. For each $i \in I$, by $s_i \sqsubseteq_{\mathsf{f}} \Theta_i$, there exists $\pi_i'' \in \Pi_{\mathtt{II}}^{\mathcal{G},1}$ satisfying $\overline{\delta}(s_i, \pi_1, \pi_i'') \overline{\sqsubseteq_{\mathsf{f}}} \overline{\delta}(\Theta_i, \pi_2, \pi_2')$. Then we have $\sum_{i \in I} p_i \cdot \overline{\delta}(s_i, \pi_1, \pi_i') \overline{\sqsubseteq_{\mathsf{f}}} \sum_{i \in I} p_i \cdot \overline{\delta}(\Theta_i, \pi_2, \pi_2')$, by Lemma 7. The required (mixed) $\mathtt{II}$-strategy $\pi_1'$ is defined by $\pi_1'(s) = \sum_{i \in I} p_i \frac{\overline{s_i}(s)}{\Delta(s)} \pi_i''(s)$ for all $s$, and the result follows from Lemma 13. $\square$

Consequently, we are able to show that lifted probabilistic alternating forward simulations are transitive.

**Corollary 1.** *(Transitivity of alternating forward simulation) Let $\sqsubseteq_f$ be a probabilistic alternating forward I-simulation, then $\Delta_1 \overline{\sqsubseteq_f} \Delta_2$ and $\Delta_2 \overline{\sqsubseteq_f} \Delta_3$ implies $\Delta_1 \overline{\sqsubseteq_f} \Delta_3$.*

## 6 Forward I-Simulation is Sound for I-PATL

This section establishes the main result of the paper: a relationship between probabilistic forward I-simulation and I-PATL formulas. Recall that a I-PATL formula has only strategy modalities $\langle\!\langle \mathtt{I} \rangle\!\rangle$ and $\langle\!\langle \emptyset \rangle\!\rangle$, and negations are only allowed to appear immediately before the propositions. For readability we write $\langle\!\langle \mathtt{I} \rangle\!\rangle$ for $\langle\!\langle \{\mathtt{I}\} \rangle\!\rangle$. Let $\mathcal{G}$ and $\mathcal{G}'$ be two PGSs, $\Delta \in \mathcal{D}(S)$ and $\Delta' \in \mathcal{D}(S')$ such that $\Delta \overline{\sqsubseteq_f} \Delta'$ by a probabilistic alternating forward I-simulation. We need to show that $\Delta \models \phi$ implies $\Delta' \models \phi$ for all I-PATL formula $\phi$.

Our proof relies on the existence of player II's optimal strategies for path formulas as winning objectives (as shown in Sect. 4). Suppose $\pi_1$ is a I strategy that enforces $\phi$, we construct another I strategy $\pi_1'$ that simulates $\pi$ all along the way, in the sense that provided the optimal II strategy $\pi_2'$ there exists another II strategy $\pi_2$ such that the probabilistic execution $\mathcal{E}(\mathcal{G}, \langle \pi_1, \pi_2 \rangle, \Delta)$ will be "simulated" by the probabilistic execution $\mathcal{E}(\mathcal{G}', \langle \pi_1', \pi_2' \rangle, \Delta')$. Since $\pi_1$ enforces $\phi$, the $\mathcal{E}(\mathcal{G}, \langle \pi_1, \pi_2 \rangle, \Delta)$ satisfies $\phi$, we show that it is also the case of $\mathcal{E}(\mathcal{G}', \langle \pi_1', \pi_2' \rangle, \Delta')$.

Let $\mathcal{E} = \langle E, \Delta, \mathcal{L}^{\mathcal{E}}, \delta^{\mathcal{E}} \rangle$ and $\mathcal{E}' = \langle E', \Delta', \mathcal{L}^{\mathcal{E}'}, \delta^{E'} \rangle$ be probabilistic executions of $\mathcal{G}$ and $\mathcal{G}'$, respectively. Also let $\sqsubseteq_f \subseteq S \times \mathcal{D}(S')$ be a probabilistic alternating forward I-simulation. We say the pair $(\mathcal{E}, \mathcal{E}')$ is an *instance of simulation*, by writing $\mathcal{E} \sqsubseteq \mathcal{E}'$, if there exists a (simulation) relation $\sqsubseteq' \subseteq E \times \mathcal{D}(E')$, such that

- $\Delta \overline{\sqsubseteq}' \Delta'$,
- if $e \sqsubseteq' \Theta$ then $last(e) \sqsubseteq_f last(\Theta)$,
- if $e \sqsubseteq' \Theta$ then $\delta^{\mathcal{E}}(e) \overline{\sqsubseteq} \delta^{\mathcal{E}'}(\Theta)$,

where $last(\Theta)$ is a distribution satisfying $last(\Theta)(s) = \sum_{last(e)=s} \Theta(e)$. A few properties of the relation $\sqsubseteq'$ are as follows.

**Lemma 15.** *1. $\Delta \overline{\sqsubseteq}' \Theta$ implies $\delta^{\mathcal{E}}(\Delta) \overline{\sqsubseteq}' \delta^{\mathcal{E}'}(\Theta)$.*
*2. $\Delta \overline{\sqsubseteq}' \Theta$ and $\Delta = \Delta_1 \oplus_\alpha \Delta_2$ with $\alpha \in [0,1]$, then there exist $\Theta_1, \Theta_2$ such that $\Delta_1 \overline{\sqsubseteq}' \Theta_1$, $\Delta_2 \overline{\sqsubseteq}' \Theta_2$, and $\Theta = \Theta_1 \oplus_\alpha \Theta_2$.*

A proof of part (1) is by definition of $\sqsubseteq'$ and Lemma 9, and part (2) holds by Lemma 10.

Let $\Delta$ be a state distribution of $\mathcal{G}$, $\Delta'$ be a state distribution of $\mathcal{G}'$, and $\Delta \overline{\sqsubseteq_f} \Delta'$. Suppose $\pi_1$ is a I strategy in $\mathcal{G}$ that enforces $\phi$ with probability at least $\alpha$, and $\pi_2'$ is a II strategy in $\mathcal{G}'$, step-by-step we establish a I strategy $\pi_1'$ and a II strategy $\pi_2$, so that the probabilistic executution decided by $\pi_1$ and $\pi_2$ from $\Delta$ will be simulated by the probabilistic executution decided by $\pi_1$ and $\pi_2$ from $\Delta'$.

**Lemma 16.** *Let $\mathcal{G} = \langle S, s_0, \mathcal{L}, \mathtt{Act}, \delta \rangle$ and $\mathcal{G}' = \langle S', s_0', \mathcal{L}', \mathtt{Act}', \delta' \rangle$ be two PGSs. If $\Delta \overline{\sqsubseteq_f} \Delta'$, then for all $\pi_1 \in \Pi_{\mathtt{I}}^{\mathcal{G}}$ and $\pi_2' \in \Pi_{\mathtt{II}}^{\mathcal{G}'}$, there exists $\pi_1' \in \Pi_{\mathtt{I}}^{\mathcal{G}'}$ and $\pi_2 \in \Pi_{\mathtt{II}}^{\mathcal{G}}$, such that $\mathcal{E}(\mathcal{G}, \langle \pi_1, \pi_2 \rangle, \Delta) \sqsubseteq \mathcal{E}'(\mathcal{G}', \langle \pi_1', \pi_2' \rangle, \Delta')$.*

*Proof.* We construct $\pi_1^i$ and $\pi_2^i$ as a level 1 strategies of player I and II for all $i \in \mathbb{N}$, and define $\pi_1'(\gamma \cdot s) = \pi_1^{|\gamma|+1}(s)$ for all $\gamma \in S^*$ and $s \in S$. And $\pi_2(\gamma \cdot s) = \pi_2^{|\gamma|+1}(s)$ for all $\gamma \in (S')^*$ and $s \in S'$.

Since $\Delta \overline{\sqsubseteq_f} \Delta'$, then by Lemma 14, there exists $\pi_1^1 \in \Pi_{\text{I}}^{\mathcal{G}',1}$, such that for all $\pi_2'' \in \Pi_{\text{II}}^{\mathcal{G}',1}$ there exists $\pi_2''' \in \Pi_{\text{II}}^{\mathcal{G},1}$ such that $\overline{\delta}(\Delta, \pi_1, \pi_2''') \widetilde{\sqsubseteq_f} \overline{\delta}(\Delta', \pi_1^1, \pi_2'')$. So if we take the first level of $\pi_2'$, there exists $\pi_2^1 \in \Pi_{\text{II}}^{\mathcal{G},1}$, such that $\overline{\delta}(\Delta, \pi_1, \pi_2^1) \widetilde{\sqsubseteq_f} \overline{\delta}(\Delta', \pi_1^1, \pi_2')$. We define $\Delta_2 \in \mathcal{D}(S^2)$ by $\Delta_2(s_1 s_2) = \Delta(s_1) \cdot \overline{\delta}(s_1, \pi_1, \pi_2^1)(s_2)$, and $\Delta_2' \in \mathcal{D}((S')^2)$ in a similar way. We also 'truncate' the strategy $\pi_1$ by defining $\pi_1(2) \in \Pi_{\text{I}}^{\mathcal{G},1}$ in the way that $\pi_1(2)(s) = \sum_{s' \in \lceil \Delta \rceil} \Delta(s') \cdot \pi_1(s's)$. And we define $\pi_2'(2)$ in a similar way.

Suppose we have $\Delta_n, \Delta_n' \in \mathcal{D}(S^n)$, and $\pi_1(n) \in \Pi_{\text{I}}^{\mathcal{G},1}$, and $\pi_2'(s) \in \Pi_{\text{II}}^{\mathcal{G}',1}$, in the similar way to above, we construct $\pi_1^n$ and $\pi_2^n$, such that $\overline{\delta}(\Delta, \pi_1(n), \pi_2^n) \widetilde{\sqsubseteq_f} \overline{\delta}(\Delta', \pi_1^n, \pi_2'(n))$. Then we define $\Delta_{n+1} \in \mathcal{D}(S^{n+1})$ by $\Delta_{n+1}(s_1 \ldots s_n s_{n+1}) = \Delta_n(s_1 \ldots s_n) \cdot \overline{\delta}(s_n, \pi_1(n), \pi_2^n)(s_{n+1})$, and $\Delta_{n+1}' \in \mathcal{D}((S')^{n+1})$ by $\Delta_{n+1}'(s_1 \ldots s_n s_{n+1}) = \Delta_n'(s_1 \ldots s_n) \cdot \overline{\delta}(s_n', \pi_1^n, \pi_2(n))(s_{n+1})$. We then define $\pi_1(n+1) \in \Pi_{\text{I}}^{\mathcal{G},1}$ by $\pi_1(n+1)(s) = \sum_{\gamma \in \lceil \Delta_n \rceil} \Delta_n(\gamma) \cdot \pi_1(\gamma \cdot s)$, and $\pi_2'(n+1) \in \Pi_{\text{II}}^{\mathcal{G}',1}$ by $\pi_2'(n+1)(s) = \sum_{\gamma \in \lceil \Delta_n' \rceil} \Delta_n'(\gamma) \cdot \pi_2'(\gamma \cdot s)$.

It is easily verifiable that we have established two probabilistic executions satisfying $\mathcal{E}(\mathcal{G}, \langle \pi_1, \pi_2 \rangle, \Delta) \sqsubseteq \mathcal{E}'(\mathcal{G}', \langle \pi_1', \pi_2' \rangle, \Delta')$, by taking a probabilistic alternating forward simulation as $\sqsubseteq'$. □

In order to measure the probability of a path formula to be satisfied when the strategies from both player I and player II are fixed, we define a relation $\models^{\bowtie \alpha}$ for probabilistic executions.

**Definition 3.** *Let $\mathcal{G}$ be a probabilistic game structure, $\mathcal{E}(\Delta) = \langle E, \Delta, \mathcal{L}^{\mathcal{E}}, \delta^{\mathcal{E}} \rangle$ a probabilistic execution determined by a strategy vector $\boldsymbol{\pi}_{\mathcal{E}}$, and $\psi$ a path formula, define*

$$\mathcal{E}(\Delta) \models^{\bowtie \alpha} \psi \quad \text{iff} \quad Pr_{\mathcal{E}}^{\Delta}(\{\rho \in \bigcup_{s \in \lceil \Delta \rceil} \mathcal{G}(\boldsymbol{\pi}_{\mathcal{E}}, s) \mid \rho \models \psi\}) \bowtie \alpha$$

It is conceivable that in a probabilistic execution every finite or infinite trace in $E^* \cup E^\omega$ maps to a trace in $\mathcal{G}$, in the way that $\rho = e_1 e_2 e_3 \ldots$ is a trace in $\mathcal{E}$ implies that $proj(\rho) = last(e_1) last(e_2) last(e_3) \ldots$ is a play in $\mathcal{G}$, where the function *proj* projects every finite sequence of states in $E$ into its last state in $S$. Consequently, we let $Pr_{\mathcal{E}}^{\Delta}$ be a probabilistic measure over $E^\omega$, such that for the cone sets (of finite traces), we have $Pr_{\mathcal{E}}^{\Delta}(e) = \Delta(last(e))$, and $Pr_{\mathcal{E}}^{\Delta}(\gamma \cdot e_1 \cdot e_2) = Pr_{\mathcal{E}}^{\Delta}(\gamma \cdot e_1) \cdot \delta^{\mathcal{E}}(e_1)(e_2)$, for $\gamma \in E^*$ and $e_1, e_2 \in E$. Let $\rho$ be an infinite trace in $\mathcal{E}$, we write $\rho \models \psi$ iff $proj(\rho) \models \psi$. Similarly, for a state formula $\phi$ and $e \in E$, write $e \in [\![\phi]\!]$ iff $last(e) \in [\![\phi]\!]$. In the following we study the properties of the satisfaction relation for a probabilistic execution to satisfy a I-PATL path formula by means of unfolding.

**Lemma 17.** *Let $\phi$, $\phi_1$ and $\phi_2$ be I-PATL (state) formulas, and $\bowtie \in \{>, \geq\}$ then*

1. $\mathcal{E}(\Delta) \models^{\bowtie \alpha} \bigcirc \phi$ iff there exists $\alpha' \bowtie \alpha$, such that $\delta^{\mathcal{E}}(\Delta) = \Delta_1 \oplus_{\alpha'} \Delta_2$ with $\lceil \Delta_1 \rceil \cap \lceil \Delta_2 \rceil = \emptyset$, and $\Delta_1 \models \phi$.
2. $\mathcal{E}(\Delta) \models^{\bowtie \alpha} \phi_1 U^{\leq k} \phi_2$ iff there exists a finite sequence of triples $\{\langle (\Delta_{i,0}, \alpha_{i,0}), (\Delta_{i,1}, \alpha_{i,1}), (\Delta_{i,2}, \alpha_{i,2}) \rangle\}_{0 \leq i \leq j}$ for some $j \leq k$, with $\lceil \Delta_{i,\ell} \rceil \cap \lceil \Delta_{i,\ell'} \rceil = \emptyset$ for all distinct $\ell, \ell' \in \{0, 1, 2\}$ and $0 \leq i \leq j$, such that

$$(1) \quad \sum_{i \in [0\ldots j]} \left( \alpha_{i,1} \cdot \prod_{i' \in [0\ldots i-1]} \alpha_{i',0} \right) \bowtie \alpha,$$

(2) $\Delta = \sum_{\ell \in \{0,1,2\}} \alpha_{0,\ell} \cdot \Delta_{0,\ell}$, and $\delta^{\mathcal{E}}(\Delta_{i,0}) = \sum_{\ell \in \{0,1,2\}} \alpha_{i+1,\ell} \cdot \Delta_{i+1,\ell}$ for all $0 \leq i < j$, (3) $\Delta_{i,0} \models \phi_1$ and $\Delta_{i,1} \models \phi_2$ for all $0 \leq i \leq j$.

3. $\mathcal{E}(\Delta) \models^{\bowtie \alpha} \phi_1 U \phi_2$ iff there exists a finite or infinite sequence of triples $\{\langle (\Delta_{i,0}, \alpha_{i,0}), (\Delta_{i,1}, \alpha_{i,1}), (\Delta_{i,2}, \alpha_{i,2}) \rangle\}_{0 \leq i < j}$ for some $j \in \mathbb{N}^+ \cup \{\infty\}$, with $\lceil \Delta_{i,\ell} \rceil \cap \lceil \Delta_{i,\ell'} \rceil = \emptyset$ for all distinct $\ell, \ell' \in \{0, 1, 2\}$ and $0 \leq i < j$, such that

$$(1) \quad \sum_{0 \leq i < j} \left( \alpha_{i,1} \cdot \prod_{i' \in [0\ldots i-1]} \alpha_{i',0} \right) \bowtie \alpha,$$

(2) $\Delta = \sum_{\ell \in \{0,1,2\}} \alpha_{0,\ell} \cdot \Delta_{0,\ell}$, and $\delta^{\mathcal{E}}(\Delta_{i,0}) = \sum_{\ell \in \{0,1,2\}} \alpha_{i+1,\ell} \cdot \Delta_{i+1,\ell}$ for all $0 \leq i < j$, (3) $\Delta_{i,0} \models \phi_1$ and $\Delta_{i,1} \models \phi_2$ for all $0 \leq i < j$.

For readability we leave the proof of this lemma in the appendix.

**Theorem 1.** *Let $\mathcal{G} = \langle S, s_0, \mathcal{L}, \mathtt{Act}, \delta \rangle$ and $\mathcal{G}' = \langle S', s'_0, \mathcal{L}', \mathtt{Act}', \delta' \rangle$ be two PGSs, $\sqsubseteq_{\mathsf{f}} \subseteq S \times \mathcal{D}(S')$ a probabilistic alternating forward $\mathtt{I}$-simulation. If $\Delta \overline{\sqsubseteq_{\mathsf{f}}} \Delta'$, then $\mathcal{G}, \Delta \models \phi$ implies $\mathcal{G}', \Delta' \models \phi$ for all $\phi \in \mathbb{L}_{\mathtt{I}}$.*

*Proof.* (sketch) We prove by induction on the structure of a $\mathtt{I}$-PATL formula $\phi$. Base case: suppose $\Delta \models p$, then $s \models p$ for all $s \in \lceil \Delta \rceil$. By $\Delta \overline{\sqsubseteq_{\mathsf{f}}} \Delta'$, there exists an index set $\{q_i\}_{i \in I}$ satisfying $\sum_{i \in I} q_i = 1$, $\Delta = \sum_{i \in I} q_i \overline{s_i}$, $\Delta' = \sum_{i \in I} q_i \Delta_i$, and $s_i \sqsubseteq_{\mathsf{f}} \Delta_i$. Therefore $\mathcal{L}(s_i) = \mathcal{L}'(t)$ for all $t \in \lceil \Delta_i \rceil$. So $t \models p$ for all $t \in \lceil \Delta_i \rceil$ for all $i$. Therefore $\Delta' \models p$. The case of $\neg p$ is similar.

We show the case when $\phi = \langle\!\langle \mathtt{I} \rangle\!\rangle^{>\alpha} \phi_1 U \phi_2$, and the proof methods for the other PATL path constructors are just similar. Since for all $t \in \lceil \Delta' \rceil$ there exists an optimal strategy $\pi^t$ for the winning objective $\neg \phi_1 \mathcal{R} \neg \phi_2$ by Lemma 2(1), and we combine these strategies into a single strategy $\pi'_2$ satisfying $\pi'_2(t \cdot \alpha) = \pi^t(t \cdot \alpha)$ for all $t \in \lceil \Delta' \rceil$ and $\alpha \in S^*$. Then $\pi'_2$ is optimal for $\neg \phi_1 \mathcal{R} \neg \phi_2$ on $\Delta'$. Then by Lemma 16, there exist $\pi_2 \in \Pi^{\mathcal{G}}_{\mathtt{II}}$ and $\pi'_1 \in \Pi^{\mathcal{G}'}_{\mathtt{I}}$ such that $\mathcal{E}(\mathcal{G}, \langle \pi_1, \pi_2 \rangle, \Delta) \sqsubseteq \mathcal{E}'(\mathcal{G}', \langle \pi'_1, \pi'_2 \rangle, \Delta')$. Since $\pi_1$ enforces $\phi_1 U \phi_2$ with probability greater than $\alpha$, we have $\mathcal{E}(\Delta) \models^{>\alpha} \phi_1 U \phi_2$.

Then by Lemma 17(3) there exists a finite or infinite sequence of triples $\{\langle (\Delta_{i,0}, \alpha_{i,0}), (\Delta_{i,1}, \alpha_{i,1}), (\Delta_{i,2}, \alpha_{i,2}) \rangle\}_{0 \leq i < j}$ for some $j \in \mathbb{N}^+ \cup \{\infty\}$ satisfying the properties as stated in Lemma 17(3). By repetitively applying Lemma 15 we establish another sequence of triples $\{\langle (\Delta'_{i,0}, \alpha_{i,0}), (\Delta'_{i,1}, \alpha_{i,1}), (\Delta'_{i,2}, \alpha_{i,2}) \rangle\}_{0 \leq i < j}$, such that (1) $\sum_{0 \leq i < j} (\alpha_{i,1} \cdot \prod_{i' \in [0\ldots i-1]} \alpha_{i',0}) > \alpha$, (2) $\Delta' = \sum_{\ell \in \{0,1,2\}} \alpha_{0,\ell} \cdot \Delta'_{0,\ell}$, and $\delta^{\mathcal{E}}(\Delta'_{i,0}) = \sum_{\ell \in \{0,1,2\}} \alpha_{i+1,\ell} \cdot \Delta'_{i+1,\ell}$ for all $0 \leq i < j$, (3) $\Delta_{i,0}\overline{\sqsubseteq_{\mathsf{f}}}\Delta'_{i,0}$ and $\Delta_{i,1}\overline{\sqsubseteq_{\mathsf{f}}}\Delta'_{i,1}$ for all $0 \leq i < j$. By induction hypothesis we have $\Delta'_{i,0} \models \phi_1$ and $\Delta'_{i,1} \models \phi_2$ for all $0 \leq i < j$. Therefore $\mathcal{E}(\Delta') \models^{>\alpha} \phi_1 \mathsf{U} \phi_2$ by Lemma 17(3).

Since $\pi'_2$ is an optimal strategy of $\mathtt{II}$, we have $\Delta' \models \langle\!\langle \mathtt{I} \rangle\!\rangle^{>\alpha} \phi_1 \mathsf{U} \phi_2$ by Lemma 3.

For a formula $\langle\!\langle \emptyset \rangle\!\rangle^{\bowtie \alpha} \psi$ we apply the same proof strategies as for $\langle\!\langle \mathtt{I} \rangle\!\rangle^{\bowtie \alpha} \psi$, except that player $\mathtt{I}$ does not need to enforce $\psi$ with a certain probability $\bowtie \alpha$ since every probabilistic execution generated by a pair of $\mathtt{I}$ and $\mathtt{II}$ strategies will enforce $\psi$ with that probability. □

## 7 Probabilistic Alternating Bisimulation

If a probabilistic alternating simulation is symmetric, we call it a probabilistic alternating bisimulation.

**Definition 4.** *Consider two probabilistic game structures $\mathcal{G} = \langle S, s_0, \mathcal{L}, \mathtt{Act}, \delta \rangle$ and $\mathcal{G}' = \langle S', s'_0, \mathcal{L}', \mathtt{Act}', \delta' \rangle$. A probabilistic alternating $\mathtt{I}$-bisimulation $\simeq \subseteq S \times S'$ is a symmetric relation satisfying if $s \simeq s'$, then*

- $\mathcal{L}(s) = \mathcal{L}'(s')$,
- *for all $\pi_1 \in \Pi_{\mathtt{I}}^{\mathcal{G},1}$, there exists $\pi'_1 \in \Pi_{\mathtt{I}}^{\mathcal{G}',1}$, such that for all $\pi'_2 \in \Pi_{\mathtt{II}}^{\mathcal{G}',1}$, there exists $\pi_2 \in \Pi_{\mathtt{II}}^{\mathcal{G},1}$, such that $\delta(s, \pi_1, \pi_2) \overline{\simeq} \delta'(s', \pi'_1, \pi'_2)$,*

*where $\overline{\simeq}$ is a lifting of $\simeq$ by weight functions.*

Since every probabilistic alternating $\mathtt{I}$-simulation is also a probabilistic alternating forward $\mathtt{I}$-simulation by treating the right hand side state as a point distribution (Lemma 11), the lifted probabilistic alternating $\mathtt{I}$-simulation is also a lifted probabilistic alternating forward $\mathtt{I}$-simulation. This fact extends for bisimulation. A probabilistic alternating $\mathtt{I}$-bisimulation also preserves formulas in $\mathbb{L}_{\mathtt{I}}$. Moreover we write $\mathbb{L}_{\mathtt{I}}^+$ for the set of formulas defined as follows, which allows negations to appear anywhere in a formula, and further we are able to show that probabilistic alternating bisimulation preserves all properties expressed in $\mathbb{L}_{\mathtt{I}}^+$.

$$\phi := p \mid \neg \phi \mid \phi_1 \wedge \phi_2 \mid \langle\!\langle A' \rangle\!\rangle^{\bowtie \alpha} \bigcirc \phi \mid \langle\!\langle A' \rangle\!\rangle^{\bowtie \alpha} \phi_1 \mathsf{U}^{\leq k} \phi_2 \mid \langle\!\langle A' \rangle\!\rangle^{>\alpha} \phi_1 \mathsf{U} \phi_2$$

**Theorem 2.** *Let $\mathcal{G} = \langle S, s_0, \mathcal{L}, \mathtt{Act}, \delta \rangle$ and $\mathcal{G}' = \langle S', s'_0, \mathcal{L}', \mathtt{Act}', \delta' \rangle$ be two PGSs, $\simeq \subseteq S \times S'$ is a probabilistic alternating $\mathtt{I}$-bisimulation. For all $s \in S$ and $s' \in S'$ with $s \simeq s'$ and $\phi \in \mathbb{L}_{\mathtt{I}}^+$, we have $\mathcal{G}, s \models \phi$ iff $\mathcal{G}', s' \models \phi$.*

The proof methodology basically follows that of Theorem 1, besides that whenever $\Delta \overline{\simeq} \Delta'$ and $\Delta \models \neg \phi$, we show that if there were $s' \in \lceil \Delta \rceil'$ such that $\mathcal{G}', s' \models \phi$ then we would also have $\mathcal{G}, s \models \phi$ for some $s \in \lceil \Delta \rceil$, which is a contradiction. And from that we have $\Delta' \models \neg \phi$ as well.

## 8 Conclusion and Future Work

We report our first results on probabilistic alternating simulation relations. We have introduced two notions of simulation for probabilistic game structures – probabilistic alternating simulation and probabilistic alternating forward simulation, following the seminal works of Segala and Lynch [Seg95a,SL95] on probabilistic simulation relations and the work of Alur et al. [AHKV98] on alternating refinement relations for non-probabilistic game structures. Our main effort has been devoted to a logical characterization for probabilistic alternating simulation relations, by showing that they preserve a fragment of PATL formulas.

On our way to the main result, we find that the proof strategy accommodated in [AHKV98] no longer applies, due to the failure in reconstructing a strategy from sub-strategies with the existence of probabilistic behaviors. Note that alternating simulations rely on mimicking behaviors by strategies of depth one, while enforcing a PATL property needs to fix a general strategy (of infinite depth) from one party regardless of any strategies of the other. We circumvent this problem by incorporating the results of probabilistic determinacy [Mar98] and the existence of optimal strategies [dAM04] in stochastic games.

There are several ways to proceed. We want to study the completeness of logical characterization for probabilistic alternating forward simulation. It is also of our interest to investigate the complexity for checking probabilistic alternating simulation relations by studying the results in the literature [AHKV98,BEMC00]. Our work was partially motivated by the paper [ASW09], where PATL is used to formalize a *balanced* property for a probabilistic contract signing protocol. Here, a balanced protocol means that a dishonest participant never has a strategy to unilaterally determine the outcome of the protocol. It is interesting to see how much the development of simulation relations for probabilistic game structures can help the verification of such kind of security protocols.

## References

[AHK97]    R. Alur, T. A. Henzinger, and O. Kupferman. Alternating-time temporal logic. In *Proc. 38th Annual Symposium on Foundations of Computer Science*, pages 100–109. IEEE Computer Society, 1997.

[AHK02]    R. Alur, T. A. Henzinger, and O. Kupferman. Alternating-time temporal logic. *Journal of ACM*, 49(5):672–713, 2002.

[AHKV98]   R. Alur, T. A. Henzinger, O. Kupferman, and M. Y. Vardi. Alternating refinement relations. In *Proc. 9th Conference on Concurrency Theory*, volume 1466 of *Lecture Notes in Computer Science*, pages 163–178. Springer, 1998.

[ASW09]    M. Aizatulin, H. Schnoor, and T. Wilke. Computationally sound analysis of a probabilistic contract signing protocol. In *Proc. 14th European Symposium on Research in Computer Security*, volume 5789 of *Lecture Notes in Computer Science*, pages 571–586. Springer, 2009.

[BEMC00]   C. Baier, B. Engelen, and M. E. Majster-Cederbaum. Deciding bisimilarity and similarity for probabilistic processes. *Journal of Computer and System Sciences*, 60(1):187–231, 2000.

[CdAH06] K. Chatterjee, L. de Alfaro, and T. A. Henzinger. The complexity of quantitative concurrent parity games. In *Proc. 17th Annual ACM-SIAM Symposium on Discrete Algorithm*, pages 678–687. ACM, 2006.

[CL07] T. Chen and J. Lu. Probabilistic alternating-time temporal logic and model checking algorithm. In *Proc. 4th Conference on Fuzzy Systems and Knowledge Discovery*, pages 35–39. IEEE Computer Society, 2007.

[dAHK98] L. de Alfaro, T. A. Henzinger, and O. Kupferman. Concurrent reachability games. In *Proc. 39th Annual IEEE Symposium on Foundations of Computer Science*, pages 564–575. IEEE Computer Society, 1998.

[dAM04] L. de Alfaro and R. Majumdar. Quantitative solution of omega-regular games. *Journal of Computer and System Sciences*, 68(2):374–397, 2004.

[DGJP02] J. Desharnais, V. Gupta, R. Jagadeesan, and P. Panangaden. Weak bisimulation is sound and complete for PCTL$^\star$. In *Proc. 13th Conference on Concurrency Theory*, volume 2421 of *Lecture Notes in Computer Science*, pages 355–370. Springer, 2002.

[Eme90] E. A. Emerson. Temporal and modal logic. In *Handbook of Theoretical Computer Science (B)*, pages 955–1072. MIT Press, 1990.

[Han94] H. Hansson. *Time and Probability in Formal Design of Distributed Systems*. Elsevier, 1994.

[LSV07] N. A. Lynch, R. Segala, and F. W. Vaandrager. Observing branching structure through probabilistic contexts. *SIAM Journal of Computing*, 37(4):977–1013, 2007.

[Mar98] D. A. Martin. The determinacy of Blackwell games. *Journal of Symbolic Logic*, 63(4):1565–1581, 1998.

[Mil89] R. Milner. *Communication and Concurrency*. Prentice Hall, 1989.

[PS07] A. Parma and R. Segala. Logical characterizations of bisimulations for discrete probabilistic systems. In *Proc. 10th Conference on Foundations of Software Science and Computational Structures*, volume 4423 of *Lecture Notes in Computer Science*, pages 287–301. Springer, 2007.

[RF91] T. E. S. Raghavan and J. A. Filar. Algorithms for stochastic games – A survey. *Mathematical Methods of Operations Research*, 35(6):437–472, 1991.

[Seg95a] R. Segala. A compositional trace-based semantics for probabilistic automata. In *Proc. 6th Conference on Concurrency Theory*, volume 962 of *Lecture Notes in Computer Science*, pages 234–248. Springer, 1995.

[Seg95b] R. Segala. *Modeling and Verification of Randomized Distributed Real-Time Systems*. PhD thesis, Massachusetts Institute of Technology, 1995.

[SL95] R. Segala and N. A. Lynch. Probabilistic simulations for probabilistic processes. *Nordic Journal of Computing*, 2(2):250–273, 1995.

[Tho91] W. Thomas. Automata on infinite objects. In *Handbook of Theoretical Computer Science, (Vol. B): Formal Models and Sematics*, pages 133–192. Elsevier, 1991.

[vGSS95] R. J. van Glabbeek, S. A. Smolka, and B. Steffen. Reactive, generative, and stratified models of probabilistic processes. *Information and Computation*, 121:59–80, 1995.

[vNM47] J. von Neumann and O. Morgenstern. *Theory of Games and Economic Behavior*. Princeton Unviersity Press, 1947.

## A  A proof of Lemma 2

The proof relies on the representation of a solution of an LTL path formula as a winning objective in quantitative game $\mu$-calculus [dAM04] for a two-player (I and $\text{II}$) game. Its grammar is defined as follows.

$$\phi := Q \mid x \mid \phi_1 \vee \phi_2 \mid \phi_1 \wedge \phi_2 \mid Ppre_{\text{I}}(\phi) \mid Ppre_{\text{II}}(\phi) \mid \mu x.\phi \mid \nu x.\phi$$

The semantics of such formulas map each formula into $\mathcal{F}$, the function space $S \to [0,1]$. A member $f \in \mathcal{F}$ gives an expected value $f(s)$ for player $\text{I}$ to win the game on every state $s \in S$. There is a partial order defined on $\mathcal{F}$ in the way that given two functions $f, g \in \mathcal{F}$, $f \leq g$ if $f(s) \leq g(s)$ for all $s \in S$. For $Q \subseteq S$, it represents a function that $Q(s) = 1$ if $s \in Q$ and $Q(s) = 0$ otherwise. For conjunction and disjunction, they are defined as $(f \wedge g)(s) = min\{f(s), g(s)\}$ for all $s \in S$, and $(f \vee g)(s) = max\{f(s), g(s)\}$ for all $s \in S$. The quantitative predecessor operator $Ppre_{\text{I}}$ for player $\text{I}$ and for every $f \in \mathcal{F}$ is by $Ppre_{\text{I}}(f)(s) = \bigsqcup_{\pi_1 \in \Pi_{\text{I}}} \bigsqcap_{\pi_2 \in \Pi_{\text{II}}} \sum_{s' \in \lceil \overline{\delta}(s, \langle \pi_1, \pi_2 \rangle) \rceil} \overline{\delta}(s, \langle \pi_1, \pi_2 \rangle)(s') f(s')$ for all $s \in S$. The operator $Ppre_{\text{II}}$ can be defined in a similar way. Intuitively, based on $f$, $Ppre_i(f)$ gives the maximal expectation of player $i$ on each state $s$ after one move, and the existence of such maximal strategy and values are guaranteed by the minimax theorem [vNM47]. Finally, $\mu x.\phi(x) = \bigsqcap \{f \in \mathcal{F} \mid \phi(f) \leq f\}$ and $\nu x.\phi(x) = \bigsqcup \{f \in \mathcal{F} \mid \phi(f) \geq f\}$.

The existence of the optimal strategy for player $\text{I}$ on an LTL objective can be sketched as follows.

- For $\bigcirc \phi$, we construct the optimal strategy from $Ppre_{\text{I}}(\phi)$ by solving a matrix game on each state $s \in S$ on reaching states in $[\![\phi]\!]$. In this case we only need to construct a level 1 strategy on every state, with its existence guaranteed by the minimax theorem.
- For bounded until $\phi_1 \text{U}^{\leq k} \phi_2$, we do the following construction recursively and prove the property by induction. $\phi_1 \text{U}^{\leq 0} \phi_2 \equiv [\![\phi_2]\!]$ works for every strategy in a state in $[\![\phi_2]\!]$. For $k > 0$, we interpret $\phi_1 \text{U}^{\leq k} \phi_2$ as $\phi_2 \vee (\phi_1 \wedge Ppre_{\text{I}}(\phi_1 \text{U}^{\leq k-1} \phi_2))$. Then suppose there exists an optimal strategy for $\phi_1 \text{U}^{\leq k-1} \phi_2$, we only need to prolong the optimal strategy by one additional level, based on the expected value already computed for $\phi_1 \text{U}^{\leq k-1} \phi_2$.
- The case of bounded release $\phi_1 \text{R}^{\leq k} \phi_2$ it can be shown in a similar way as the above case, by letting $\phi_1 \text{R}^{\leq 0} \phi_2$ as $[\![\phi_2]\!]$, and $\phi_1 \text{R}^{\leq k} \phi_2$ as $\phi_2 \wedge (\phi_1 \vee Ppre_{\text{I}}(\phi_1 \text{R}^{\leq k-1} \phi_2))$ for each $k > 0$.
- For unbounded release $\phi_1 \text{R} \phi_2$, our argument resembles the proof of [dAM04, Lemma 2] on safety games. The value of the game for player $\text{I}$ as the protagonist is interpreted as the function $f = \nu x. \phi_2 \wedge (\phi_1 \vee Ppre_{\text{I}}(x))$, and there exists a memoryless strategy $\pi_1 \in \Pi_{\text{I}}^{\mathcal{G},1}$ for player $\text{I}$ so that on each state $s \in S$, $\pi_1(s) \in \mathcal{D}(\text{Act}_{\text{I}})$ is the best choice (in the matrix game on $s$) player $\text{I}$ can make according to the greatest fixed point $f$, i.e., for all memoryless player $\text{II}$ strategies $\pi_2 \in \Pi_{\text{II}}^{\mathcal{G},1}$, we have $\sum_{s' \in S} \overline{\delta}(s, \pi_1, \pi_2)(s') \cdot f(s') \geq f(s)$. We show that $\pi_1$ is the strategy that guarantees $f(s)$ on each state $s$ in the general sense. Let $\pi_2 \in \Pi_{\text{II}}^{\mathcal{G}}$ be an arbitrary player $\text{II}$ strategy, and $s \in S$, we are going to show that in the probabilistic execution $\mathcal{E}(\mathcal{G}, \pi_1 \cup \pi_2, \overline{s}) = \langle E, s, \mathcal{L}, \delta \rangle$, we have $\mathcal{E}(\overline{s}) \models^{\geq f(s)} \phi_1 \text{R} \phi_2$. In order to do so, we give the following intermediate result that $\mathcal{E}(\overline{s}) \models^{\geq f(s)} \phi_1 \text{R}^n \phi_2$ for all $n \in \mathbb{N}$.

We prove this by induction on $n \in \mathbb{N}$ that $\mathcal{E}(\bar{e}) \models^{\geq f(last(e))} \phi_1 \mathtt{R}^{\leq n} \phi_2$ for all $e \in E$. By abuse of the notation we treat $\pi_1$ also as a general strategy such that $\pi_1(\gamma) = \pi_1(last(\gamma))$ for all $\gamma \in S^+$. We also write $\pi_2^e$ as the "truncated" strategy of $\pi_2$, by defining $\pi_2^e(s \cdot \gamma) = \pi_2(e \cdot \gamma)$ for all $s \in S$ and $\gamma \in S^+$. To simplify notation we write $Pr_{\mathcal{E}}^{\bar{e}}(\pi_1, \pi_2, \psi)$ for $Pr_{\mathcal{E}}^{\bar{e}}(\{\rho \in \mathcal{G}((\pi_1, \pi_2), last(e)) \mid \rho \models \psi\})$.

Base case: let $e \in E$, $n = 0$ and $\phi_1 \mathtt{R}^{\leq 0} \phi_2 \equiv \phi_2$, we have $Pr_{\mathcal{E}}^{\bar{e}}(\pi_1, \pi_2^e, \phi_1 \mathtt{R}^{\leq 0} \phi_2) = 1 \geq f(last(e)) = 1$ if $last(e) \in [\![\phi_2]\!]$, and $Pr_{\mathcal{E}}^{\bar{e}}(\pi_1, \pi_2^e, \phi_1 \mathtt{R}^{\leq 0} \phi_2) = 0 \geq f(last(e)) = 0$ otherwise.

Suppose this holds up to level $n$, we need to show the case of $n + 1$.
- If $last(e) \in [\![\neg \phi_2]\!]$, then $Pr_{\mathcal{E}}^{\bar{e}}(\pi_1, \pi_2^e, \phi_1 \mathtt{R}^{\leq n+1} \phi_2) = 0 \geq f(last(e))$.
- If $last(e) \in [\![\phi_2]\!] \cap [\![\phi_1]\!]$, then $Pr_{\mathcal{E}}^{\bar{e}}(\pi_1, \pi_2^e, \phi_1 \mathtt{R}^{\leq n+1} \phi_2) = 1 \geq f(last(e))$.
- If $last(e) \in [\![\phi_2]\!] \cap [\![\neg \phi_1]\!]$, then $Pr_{\mathcal{E}}^{\bar{e}}(\pi_1, \pi_2^e, \phi_1 \mathtt{R}^{\leq n+1} \phi_2) = \sum_{e' \in \lceil \delta(e) \rceil} \delta(e)(e') \cdot Pr_{\mathcal{E}}^{\bar{e'}}(\pi_1, \pi_2^{e'}, \phi_1 \mathtt{R}^{\leq n} \phi_2)$. By I.H., $Pr_{\mathcal{E}}^{\bar{e'}}(\pi_1, \pi_2^{e'}, \phi_1 \mathtt{R}^{\leq n} \phi_2) \geq f(last(e'))$, we have $Pr_{\mathcal{E}}^{\bar{e}}(\pi_1, \pi_2^e, \phi_1 \mathtt{R}^{\leq n+1} \phi_2) \geq \sum_{e' \in \lceil \delta(e) \rceil} \delta(e)(e') \cdot f(last(e'))$.

  By definition we have $f(last(e))$
  $= \prod_{\pi' \in \Pi_{\mathtt{II}}^{\mathcal{G}, 1}} \sum_{s' \in \bar{\delta}(last(e), \pi_1, \pi')} \bar{\delta}(last(e), \pi_1, \pi')(s') f(s')$
  $\leq \sum_{s' \in \bar{\delta}(last(e), \pi_1, \pi_2)} \bar{\delta}(last(e), \pi_1, \pi_2)(s') f(s')$
  $= \sum_{e' \in \lceil \delta(e) \rceil} \delta(e)(e') \cdot f(last(e'))$.

  The last equivalence is by definition of $\mathcal{E}(\mathcal{G}, \pi_1 \cup \pi_2, \bar{s})$. The result immediately follows.

$Pr_{\mathcal{E}}^{\bar{s}}(\pi_1, \pi_2, \phi_1 \mathtt{R} \phi_2) = \lim_{n \to \infty} Pr_{\mathcal{E}}^{\bar{s}}(\pi_1, \pi_2, \phi_1 \mathtt{R}^n \phi_2)$, we have $Pr_{\mathcal{E}}^{\bar{s}}(\pi_1, \pi_2, \phi_1 \mathtt{R} \phi_2) \geq f(s)$. Since $\pi_2$ is arbitrarily chosen, we have $\pi_1$ is a player $\mathtt{I}$ strategy that enforces $\phi_1 \mathtt{R} \phi_2$ with probability at least $f(s)$. Moreover, since $s$ is arbitrarily chosen, $\pi_1$ is optimal on $s$ for all $s \in S$.

The existence of $\epsilon$-optimal strategies for $\phi_1 \mathtt{U} \phi_2$ for all $\epsilon > 0$ is guaranteed by the existence of $\epsilon$-optimal strategies for all $\omega$-regular winning objectives, as shown in [dAM04].

## B  A proof sketch of Lemma 17

1. Define the set $\{s \in \lceil \delta^{\mathcal{E}}(\Delta) \rceil \mid s \models \phi\}$. Now it is obvious that for all infinite run $\rho \in E^\omega$, $\rho \models \bigcirc \phi$ iff $\rho(1) \models \phi$. The result immediately follows.
2. The proof this item is similar to the below in its finite case.
3. Given the probabilistic execution $\mathcal{E}(\Delta)$, intuitively, each $e \in E$ represents a finite run within $\mathcal{E}(\Delta)$. We construct as follows a *maximal* sequence of triples $\{\langle (\Delta_{i,0}, \alpha_{i,0}), (\Delta_{i,1}, \alpha_{i,1}), (\Delta_{i,2}, \alpha_{i,2}) \rangle\}$ for $i \in \mathbb{N}$.

   For all $e \in E$, we define $E_{i,1} = \{e \in E \mid e = s_1 s_2 \ldots s_i, s_i \models \phi_2, s_j \models \phi_1 \text{ for all } j < i\}$, and $E_{i,0} = \{e \in E \mid e = s_1 s_2 \ldots s_i, s_j \models \phi_1 \text{ for all } j \leq i\}$. Intuitively, $E_{i,1}$ are the prefixes of those runs that satisfy $\phi_1 \mathtt{U} \phi_2$, and $E_{i,1}$ are the prefixes of the runs that might satisfy $\phi_1 \mathtt{U} \phi_2$.

   Further we define $\Delta_0 = \Delta$ and $E_{0,2} = \lceil \Delta_0 \rceil \setminus (E_{0,0} \cup E_{0,1})$, and $\alpha_{0,\ell} = \Delta_0(E_{0,\ell})$ for $\ell \in \{0, 1, 2\}$. Then for each $i \in \mathbb{N}$, recursively define $\Delta_{i+1} = \delta(\Delta_{i,0})$, $E_{i+1,2} = \lceil \Delta_{i+1} \rceil \setminus (E_{i+1,0} \cup E_{i+1,1})$, $\alpha_{i+1,\ell} = \Delta_{i+1}(E_{i+1,\ell})$ for all

$\ell \in \{0, 1, 2\}$. Consequently, we have $\Delta_{i,\ell}(e) = \Delta_i(e)/\alpha_{i,\ell}$ if $e \in E_{i,\ell}$ and $0$ otherwise, for all $i \in \mathbb{N}$ and $\ell \in \{0, 1, 2\}$, provided $\alpha_{i,\ell} \neq 0$. If $\alpha_{i,\ell} = 0$ we let $\Delta_{i,\ell}$ be empty, i.e., it assigns every $e \in E$ to $0$.

It is easily verifiable that $\lceil \Delta_{i,\ell} \rceil \cap \lceil \Delta_{i,\ell'} \rceil = \emptyset$ for all distinct $\ell, \ell' \in \{0, 1, 2\}$, since $E_{i,\ell} \cap E_{i,\ell'} = \emptyset$. Also for all $i \in \mathbb{N}$ we have (2) $\delta^{\mathcal{E}}(\Delta_{i,0}) = \sum_{\ell \in \{0,1,2\}} \alpha_{i+1,\ell} \cdot \Delta_{i+1,\ell}$, and (3) $\Delta_{i,0} \models \phi_1$ and $\Delta_{i,2} \models \phi_2$ for all $i$.

For every infinite run $\rho \in E^\omega$, we have $\rho \models \phi_1 \mathtt{U} \phi_2$ iff there exists a prefix $e \in E_{i,1}$ for some $i \in \mathbb{N}$. Therefore we have $Pr_{\mathcal{E}}^\Delta(\{\rho \in E^\omega \mid \rho \models \phi_1 \mathtt{U} \phi_2\}) \bowtie \alpha$ iff $\sum_{i \in \mathbb{N}}(\alpha_{i,1} \cdot \prod_{0 \leq j < i} \alpha_{j,0})$ for $\bowtie \in \{>, \geq\}$, which is (1). That is, the collection of infinite traces satisfying $\phi_1 \mathtt{U} \phi_2$ are those with prefix $e_0 e_1 \ldots e_i$ with $e_i \models \phi_2$ and $e_j \models \phi_1$ for all $0 \leq j < i$. Therefore, suppose $\mathcal{E}(\Delta) \models^\bowtie \phi_1 \mathtt{U} \phi_2$, we have that the above sequence of triples $\{\langle (\Delta_{i,0}, \alpha_{i,0}), (\Delta_{i,1}, \alpha_{i,1}), (\Delta_{i,2}, \alpha_{i,2}) \rangle\}_{i \in \mathbb{N}}$ satisfies the required conditions (1), (2) and (3).

Suppose there exists a sequence of triples satisfying (1), (2) and (3) with respect to $\bowtie \alpha$, due to a similar way of reasoning, we already collected enough infinite runs that satisfy $\phi_1 \mathtt{U} \phi_2$ with probability $\bowtie \alpha$ for $\bowtie \in \{>, \geq\}$.